%% file: SUSYnnpdf.tex
\documentclass[11pt,a4paper]{article}

\usepackage{graphicx}
\usepackage{float}
\usepackage{afterpage}
\usepackage{epsfig,cite}
\usepackage{amssymb}
\usepackage{amsmath}
\usepackage{dsfont}
\usepackage{multirow}
\usepackage{url}

\newcommand{\be}{\begin{equation}}
\newcommand{\ee}{\end{equation}}
\newcommand{\bea}{\begin{eqnarray}}
\newcommand{\eea}{\end{eqnarray}}
\newcommand{\bi}{\begin{itemize}}
\newcommand{\ei}{\end{itemize}}
\newcommand{\ben}{\begin{enumerate}}
\newcommand{\een}{\end{enumerate}}

\def\frac#1#2{{{#1}\over {#2}}}
\def\gsim{\mathrel{\rlap{\lower4pt\hbox{\hskip1pt$\sim$}}
    \raise1pt\hbox{$>$}}}         
\def\lsim{\mathrel{\rlap{\lower4pt\hbox{\hskip1pt$\sim$}}
    \raise1pt\hbox{$<$}}}         

\newcommand{\draft}[1]{}

\def\beq{\begin{equation}}  
\def\eeq{\end{equation}}

\def \n0{N_j^{(0)}}

\def\lapprox{\lower .7ex\hbox{$\;\stackrel{\textstyle <}{\sim}\;$}}
\def\gapprox{\lower .7ex\hbox{$\;\stackrel{\textstyle >}{\sim}\;$}}

\begin{document}
\vspace{-2.0cm}
\begin{flushright}
  OUTP-15-18P \\ MS-TP-15-16\\ TTK-15-26\\ ITP-UU-15/13 \\
Nikhef/2015-032
\end{flushright}

\begin{center}
  {\Large \bf NLO+NLL squark and gluino production cross-sections\\[0.26em] with
    threshold-improved parton distributions}\\
    \vspace{.7cm}
  Wim Beenakker$^1$,
Christoph Borschensky$^2$,
Michael Kr\"amer$^3$,
Anna Kulesza$^2$,\\
Eric Laenen$^4$,
Simone Marzani$^5$ and
Juan~Rojo$^6$

\vspace{.3cm}
 {\it
~$^1$ Theoretical High Energy Physics, IMAPP, \\Faculty of Science, Mailbox 79, Radboud University\\
Nijmegen, P.O. Box 9010, NL-6500 GL Nijmegen,
Institute of Physics,\\ University of Amsterdam, Amsterdam, The Netherlands\\
~$^2$ Institute for Theoretical Physics, WWU M\"unster, D-48149 M\"unster, Germany\\
~$^3$ Institute for Theoretical Particle Physics and Cosmology, \\RWTH Aachen University D-52056, Aachen, Germany \\
~$^4$ ITFA, University of Amsterdam, Science Park 904, 1018 XE, Amsterdam,\\
ITF, Utrecht University, Leuvenlaan 4, 3584 CE Utrecht,\\
Nikhef Theory Group, Science Park 105, 1098 XG Amsterdam, The Netherlands\\
~$^5$ Department of Physics, University at Buffalo,\\
The State University of New York, Buffalo, NY 14260-1500, USA\\
~$^6$ Rudolf Peierls Centre for Theoretical Physics, 1 Keble Road,\\
University of Oxford, OX1 3NP Oxford, UK\\
}
\end{center}   

\vspace{0.1cm}

\begin{center}
{\bf \large Abstract}
\end{center}

We present updated predictions for the cross-sections for pair
production of squarks and gluinos at the LHC Run II.
First of all, we update the calculations
based on NLO+NLL partonic cross-sections
by using the NNPDF3.0NLO global analysis.
This study includes a full characterization of
theoretical uncertainties from higher
orders, PDFs and the strong coupling.
Then, we explore the implications for this calculation
of the recent NNPDF3.0 PDFs with NLO+NLL threshold resummation.
We find that the shift in the results
induced by the threshold-improved PDFs
is within the total theory uncertainty band of the 
calculation based on NLO PDFs.
However, we also observe that the central values of the NLO+NLL
cross-sections are modified both in a qualitative and a quantitative way,
illustrating the relevance and impact
of using threshold-improved PDFs
together with resummed partonic cross-sections.
The updated NLO+NLL cross-sections based on
NNPDF3.0NLO are publicly available
in the {\tt NLL-fast} format,
and should
be an important ingredient for the interpretation of the searches
for supersymmetric
particles at Run II.

\clearpage

\input{sec-introduction}

\input{sec-nnpdf30}

\input{sec-resummation}

\input{sec-summary}

\appendix
\input{sec-appendix}

\input{SUSYnnpdf.bbl}

\end{document}

%% file: sec-introduction.tex
\section{Introduction and motivation}
\label{sec:introduction}

Supersymmetry (SUSY)~\cite{Wess:1973kz,Wess:1974tw,Fayet:1976et,Farrar:1978xj,Sohnius:1985qm,Martin:1997ns} is one of the
better motivated extensions of the Standard Model (SM).
In addition to offering a natural solution to the hierarchy problem, it provides
a number of dark-matter candidates and leads to the unification of gauge couplings at high scales.
While so far searches for supersymmetry at the LHC 8 TeV have returned null
results~\cite{Aad:2015iea,Aad:2015baa,Khachatryan:2015lwa},
the recently started Run II of the LHC, with its increase in center-of-mass energy
and luminosity, opens a wide new region of the SUSY parameter space
for scrutiny, extending in particular
into the high-mass
regime
up to 2.5 TeV particles~\cite{Gershtein:2013iqa}.
For this reason, it is crucial to provide precise theory predictions for
the cross-sections of supersymmetric particle pair production at 13 TeV with
a robust estimate of the associated theory uncertainties.

The next-to-leading order (NLO) QCD corrections to the
total cross-sections for squark and gluino pair production
were first computed in~\cite{Beenakker:1994an,Beenakker:1995fp,Beenakker:1996ch}.
Recently, a considerable effort has been invested in automating these calculations~\cite{GoncalvesNetto:2012yt, Goncalves:2014axa,Degrande:2014sta}, including the decays~\cite{Hollik:2012rc} and matching the NLO corrections with a parton shower~\cite{Gavin:2013kga,Gavin:2014yga}.
Furthermore, NLO electroweak corrections have been also calculated ~\cite{Hollik:2007wf,Bornhauser:2007bf,Hollik:2008yi,Hollik:2008vm,Mirabella:2009ap,Bornhauser:2009ru,Germer:2010vn,Germer:2014jpa,Hollik:2015lha}.
A significant contribution to the QCD NLO corrections originates from soft-gluon emissions which dominate the region near the production threshold.
In this region, the partonic center-of-mass energy $\hat{s}$
is close to the kinematic restriction for the on-shell production of these particles, i.e. $\hat{s} \geq 4m^2$, with $m$
being the average mass of the two produced particles.
Soft-gluon corrections to squark and gluino pair production can be taken into account to all orders in perturbation theory either using threshold resummation techniques in Mellin space~\cite{Bonciani:1998vc,Contopanagos:1996nh,Kidonakis:1998bk,Kidonakis:1998nf} or in the framework of effective theories~\cite{Beneke:2009rj, Beneke:2010da}.

Resummation of threshold corrections at the next-to-leading logarithmic (NLL) accuracy has been performed for all processes contributing to squark and gluino production, resulting in matched NLO+NLL predictions~\cite{Kulesza:2008jb,Kulesza:2009kq,Beenakker:2009ha,Beenakker:2010nq,Beenakker:2011fu,Kramer:2012bx,Borschensky:2014cia}. Additionally, Coulomb corrections have been resummed simultaneously with soft gluon corrections in~\cite{Beneke:2010da,Falgari:2012hx}.
The progress towards resummation at next-to-next-to-leading logarithmic (NNLL)
accuracy and towards
the corresponding LHC predictions has been reported in~\cite{Beenakker:2011sf,Beenakker:2011gt,Beenakker:2013mva,Beenakker:2014sma,Beneke:2014wda,Broggio:2013uba, Broggio:2013cia,Pfoh:2013edr,Langenfeld:2009eg,Langenfeld:2012ti}.
Finite width and bound state effects have been studied for squark and gluino production processes in~\cite{Falgari:2012sq,Kauth:2009ud,Hagiwara:2009hq,Kauth:2011vg,Kauth:2011bz, Kim:2014yaa}.
Resummed calculations for sleptons and gauginos are also available
at NLO+NLL accuracy~\cite{Fuks:2013lya,Fuks:2013vua,Bozzi:2007tea}.

A potential source of theory uncertainty in
these calculations arises from the fact that they are based
on parton distribution functions (PDFs)
extracted from experimental data
using partonic cross-sections computed at fixed-order, NLO accuracy.
Ideally, one should use threshold-improved PDFs determined
using the same NLO+NLL theory as that used in the computation
of the supersymmetric partonic
cross-sections.
Indeed, from general considerations, one expects that
resummed
PDFs should partially cancel the typical enhancements found in NLO+NLL
partonic cross-sections.

Therefore, it is
important to ascertain the effects of using
threshold-resummed PDFs and
quantify their phenomenological
relevance for NLO+NLL SUSY cross-sections.
From the practical point of view, it is essential
to  determine if the shift induced by the threshold-improved
PDFs lies within the
estimated theory uncertainty bands, the latter
obtained from cross-sections based on fixed-order NLO PDFs.
If such a shift is larger than these uncertainty bands,
it would affect the existing SUSY exclusion limits drawn
from the LHC measurements.
Until recently, no threshold-improved PDF sets
were available.

With the motivation of being able to perform, for the
first time NLO+NLL and NNLO+NNLL threshold-resummed
calculations together with resummed
PDFs,
variants of the NNPDF3.0 analysis~\cite{Ball:2014uwa}
have now become available~\cite{Bonvini:2015ira}.
These resummed PDFs differ from their fixed-order counterparts due to
the different theory inputs used in the two cases.
While in the $\overline{\rm MS}$ scheme the DGLAP evolution kernels are unaffected by threshold resummation~\cite{Albino:2000cp}, the partonic cross-sections for the processes that enter the PDF fit  are
modified (for example for deep-inelastic
structure functions and Drell-Yan rapidity distributions).
Therefore, these differences in the partonic cross-sections used in the fit (NLO+NLL in the former case, NLO in the latter) translate into differences
in the extracted PDFs at the input fitting scale $Q_0$.
Once these boundary conditions are determined from the fit, standard DGLAP evolution can be used to evolve both the fixed-order and resummed PDFs for any other value of $Q^2\ge Q_0^2$.

A limitation of these threshold-improved NNPDF sets is that
threshold-resummed calculations are not readily available
for all the processes
included in the NNPDF3.0 global analysis,
in particular for inclusive jets
and charged current Drell-Yan production.
This implies that the resummed PDFs of~\cite{Bonvini:2015ira}
 are based on a more limited dataset than
 the global fit, and thus are affected by somewhat larger uncertainties,
 in particular
for the gluon PDF at large $x$.
Therefore, it is necessary to devise a prescription to
combine the cross-sections
from the global, fixed-order,
PDF fit,  with those of the resummed PDF fit based on the reduced dataset.

The first  motivation of the current work is to update the NLO+NLL
calculations of squark and gluino
pair production at the LHC 13 TeV from
Refs.~\cite{Borschensky:2014cia,Kramer:2012bx}
with the NNPDF3.0 NLO global PDF set~\cite{Ball:2014uwa}.
While previous NLO+NLL predictions were obtained using older global fits,
CTEQ6.6~\cite{Nadolsky:2008zw} and MSTW08~\cite{Martin:2009iq},
the new results
using NNPDF3.0 should provide a more accurate estimate of theory uncertainties
especially in the high-mass
region, where the flexible NNPDF parametrization~\cite{Ball:2008by,Ball:2010de,DelDebbio:2007ee,Ball:2011uy,Ball:2009mk} minimizes the possible introduction of theory biases that might arise in
fits with fixed functional forms.
The NNPDF3.0 analysis also
includes a variety of recent measurements, in particular
from the LHC Run I, which provide
better control on the large-$x$ PDF uncertainties.
Moreover,
the NNPDF3.0 fit is based on an extensive improvement of the positivity constraints to ensure
that, even if PDFs might become negative, physical cross-sections should
be positive, resulting in
more reliable PDFs in the very large-$x$ region as compared to the previous NNPDF2.3 set~\cite{Ball:2012cx}.

The updated NLO+NLL squark and gluino cross-sections
computed with NNPDF3.0 NLO
are available
in the {\tt NLL-fast} format
of fast interpolation tables.
They include a 
complete
characterization of theoretical uncertainties from PDFs,
higher orders and the strong coupling $\alpha_s$.
The flexible {\tt NLL-fast} format ensures that the
results of this work can be readily
used by the ATLAS and CMS collaborations for their
interpretation of their SUSY searches.

Subsequently, we present an exploration of the implications of the NNPDF
threshold-improved PDFs on the
NLO+NLL calculations of supersymmetric particle pair production.
As we will show, the result of the calculation where the
effects of threshold resummation are included
both at the level of PDFs and at the level of partonic cross-sections can differ substantially from
the calculation where resummation is only applied to partonic cross-sections.
Reassuringly, the cross-sections obtained with the
threshold-improved PDFs are
contained within the theory uncertainty band (including higher-orders and
PDF uncertainties) of the traditional calculation based on the fixed-order NNPDF3.0 NLO PDF set.
This result constitutes a non-trivial
validation
of the robustness of current estimates of the total theory
uncertainty
in SUSY production cross-sections.
While these differences, therefore,
do not affect present
exclusion limits, they would become crucial in the case
of the discovery of supersymmetric particles, since they would affect the determination of their
properties from the LHC data, in the same way as in the extraction
of Higgs couplings and branching fractions.

The outline of this paper is as follows.
In Sect.~\ref{sec-nnpdf30} we present the
update of the NLO+NLL cross-sections for squark
and gluino pair production at 13 TeV based on the NNPDF3.0 NLO global fit.
In Sect.~\ref{sec-resummation} we quantify the impact
of threshold-resummed PDFs for the NLO+NLL
supersymmetric cross-sections.
In Sect.~\ref{sec:summary} we conclude, outline possible future developments
and present the delivery of the results of this work.
Appendix~\ref{sec:tresh-resumm-pdfs} discusses
an alternative possibility to incorporate threshold resummation
effects into the PDFs.


%% file: sec-nnpdf30.tex
\section{NLO+NLL cross-sections at 13 TeV with NNPDF3.0NLO}
\label{sec-nnpdf30}

In this section we update the NLO+NLL predictions for
squark and gluino pair production cross-sections at
13 TeV from
Refs.~\cite{Borschensky:2014cia,Kramer:2012bx} using now the
recent NNPDF3.0 NLO global analysis as an input.
We focus 
on the Minimal Supersymmetric extension of the Standard Model (MSSM) with R-parity conservation, where
SUSY particles are always pair produced.
At a hadron collider such as the LHC,
 the most copiously produced SUSY particles are expected to be the
 strongly interacting partners of quarks and antiquarks, the squarks $\tilde{q}$ and anti-squarks $\tilde{q}^*$,
 and of the gluons, the gluinos $\tilde{g}$.
 Therefore, in this work we will compute total cross-sections for the
 following processes:
 \be
 pp \to \tilde{q}\tilde{q} \,, \tilde{q}\tilde{q}^*, \tilde{q}\tilde{g},
 \tilde{g}\tilde{g}+X \, ,
\label{eq:processes}
 \ee
 where by $\tilde{q}(\tilde{q}^*)$ we denote the partners of the massless
 (anti-)quarks only.

 In the present calculations, all flavours of final-state squarks are included and summed over, except top squarks, due to the large mixing effects and the mass splitting in the stop sector.
 We sum over all light-flavour squarks with both chiralities ($\tilde{q}_{L}$ and~$\tilde{q}_{R}$), which are taken to be mass-degenerate.
The QCD coupling $\alpha_{\rm s}(Q)$ and the parton distribution functions at NLO are defined in the $\overline{\rm MS}$ scheme with five active flavours. The renormalization and factorization scales are taken to be equal $\mu=\mu_R=\mu_F$ and a top-quark mass of $m_t=173.21$~GeV is used~\cite{Agashe:2014kda}.
The NLO parts of the NLO+NLL predictions are obtained using the public code {\tt Prospino}~\cite{Beenakker:1996ch} while the resummed parts  are computed as in Ref.~\cite{Beenakker:2009ha}.
We refer the reader to~\cite{Beenakker:1996ch,Kulesza:2009kq, Beenakker:2009ha} for a more thorough discussion of the theory behind the NLO+NLL calculations.
For simplicity, in this work we will present results
assuming equal squark and gluino masses $m \equiv m_{\tilde{q}}=m_{\tilde{g}}$.
In our baseline calculations the factorization and renormalization scales are set equal to
the sparticle masses, $\mu=m$.

	\begin{figure}[t]
		\centering
		\includegraphics[width=.49\textwidth]{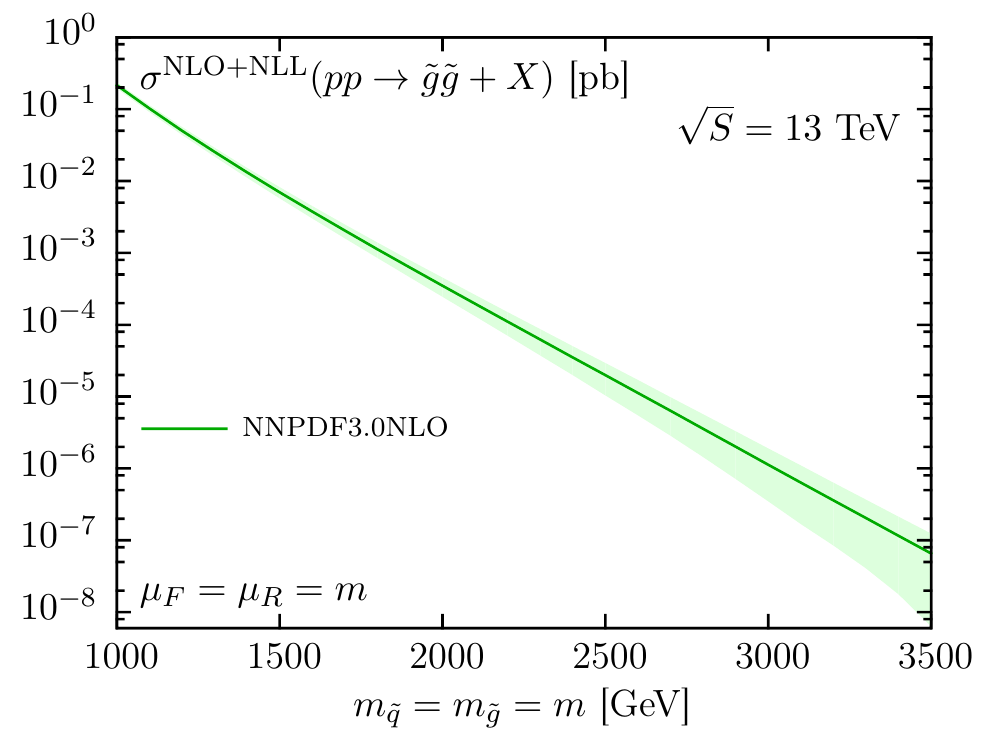}\hfill\includegraphics[width=.49\textwidth]{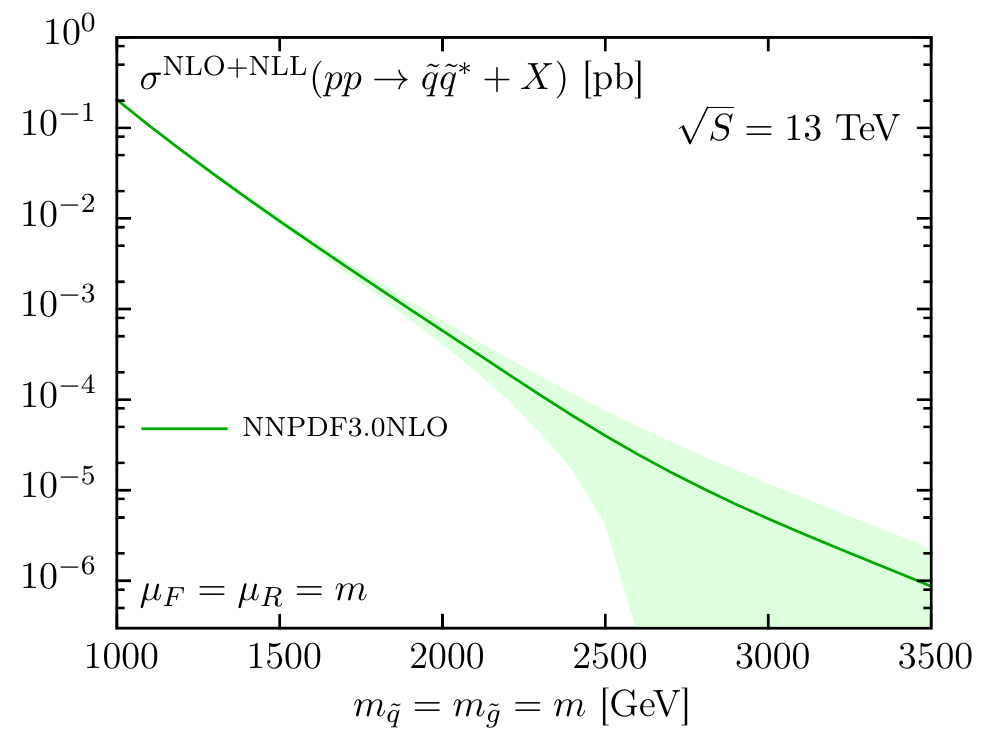}\\
		\includegraphics[width=.49\textwidth]{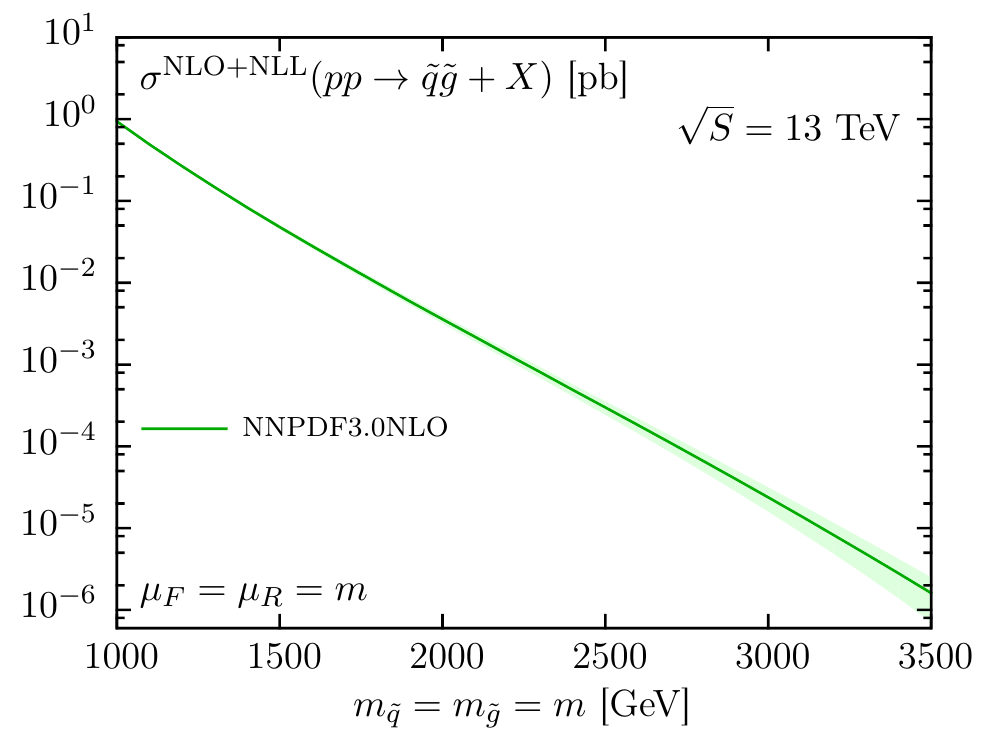}\hfill\includegraphics[width=.49\textwidth]{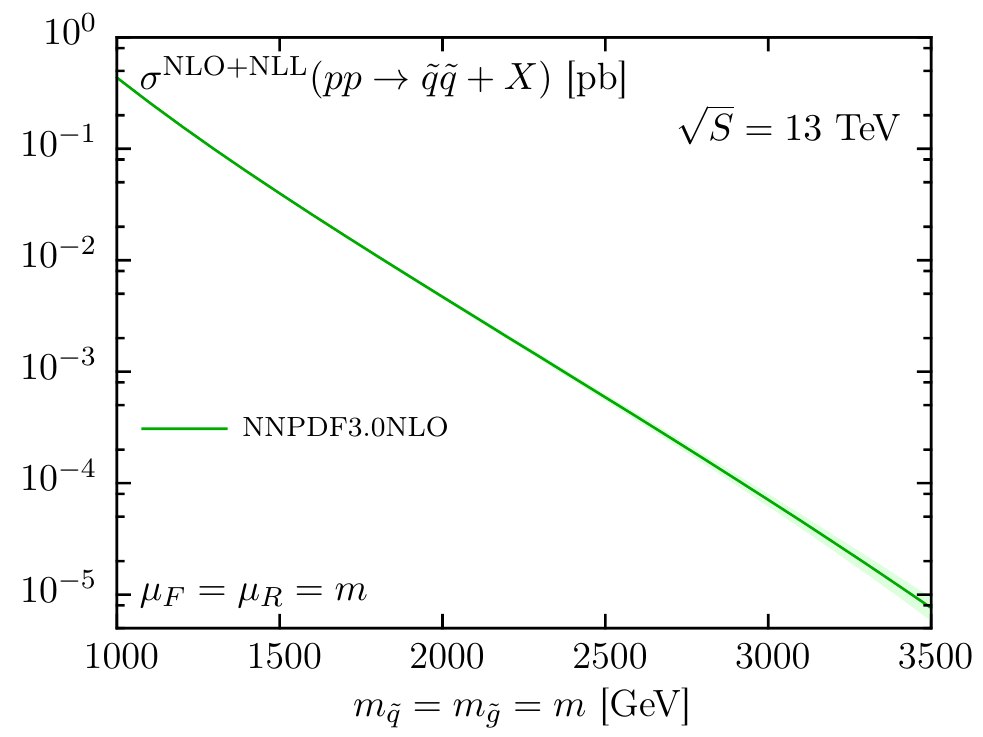}
    \caption{\small The NLO+NLL total cross-sections for the
      pair production of squarks and gluinos at the LHC $\sqrt{S}=13$ TeV
      in different
channels, obtained using the NNPDF3.0 NLO set.
Results are shown as a function of the sparticle mass $m$ and include the one-sigma PDF uncertainties.
  \label{fig:xsec1}
    }
	\end{figure}

        In Fig.~\ref{fig:xsec1} we show the NLO+NLL total cross-sections
obtained using the NNPDF3.0 NLO global fit
        for the four processes listed in
Eq.~(\ref{eq:processes}).
We present our results, as will be done
in the rest of this work,
for sparticle masses in the range
  between 1 TeV and 3.5 TeV.
  The choice of the lower range is motivated by the upper limits of existing searches
  of squarks and gluinos by ATLAS~\cite{Aad:2015iea,Aad:2015baa} and CMS~\cite{Khachatryan:2015lwa} at
  8 TeV, while the upper value coincides with the largest
  sparticle masses that can be probed at the LHC, including
  its future high-luminosity upgrade~\cite{CMS:2013xfa}.
  From these absolute cross-section plots we already see
  that the PDF uncertainties blow up at large masses,
  reflecting the lack of experimental constraints on the large-$x$ PDFs.

  It is important to point out that for some specific final states, in particular
  those driven by the $q\bar{q}$ initial state,
  for large values of the sparticle
  mass $m$, roughly for $m\gsim 2.5$ TeV, the cross-section computed with some
  replicas of the NNPDF3.0 NLO set might become negative.
  As discussed in~\cite{Ball:2014uwa}, in NNPDF3.0 no {\it ad-hoc}
  conditions are imposed
  on the shape nor on the positivity of PDFs in the large-$x$ region (to avoid
  any theory bias and the corresponding underestimating of PDF uncertainties), while
  at the same time one requires the positivity of a number of physical cross-sections,
  such as deep-inelastic structure functions, Drell-Yan rapidity distributions and
  Higgs production in gluon fusion.

  However, it is technically impossible to constrain all possible cross-sections to be positive
  during the fit, and therefore, when for specific processes in extreme regions
  of the phase space a cross-section computed with a NNPDF replica becomes negative,
  the correct prescription is to set it to zero before evaluating the PDF
  uncertainty.
  For the cross-sections of Fig.~\ref{fig:xsec1}, as well as in the rest of the
  calculations presented in this paper,  negative cross-sections are always set to zero.
  We have verified that this has only appreciable effects at very large masses,
  where PDF uncertainties are huge anyway, and moreover,
  that the shift in central values
  induced by this prescription is always negligible as compared to
  the intrinsic PDF
  uncertainty.

\begin{figure}[t]
		\centering
		\includegraphics[width=.49\textwidth]{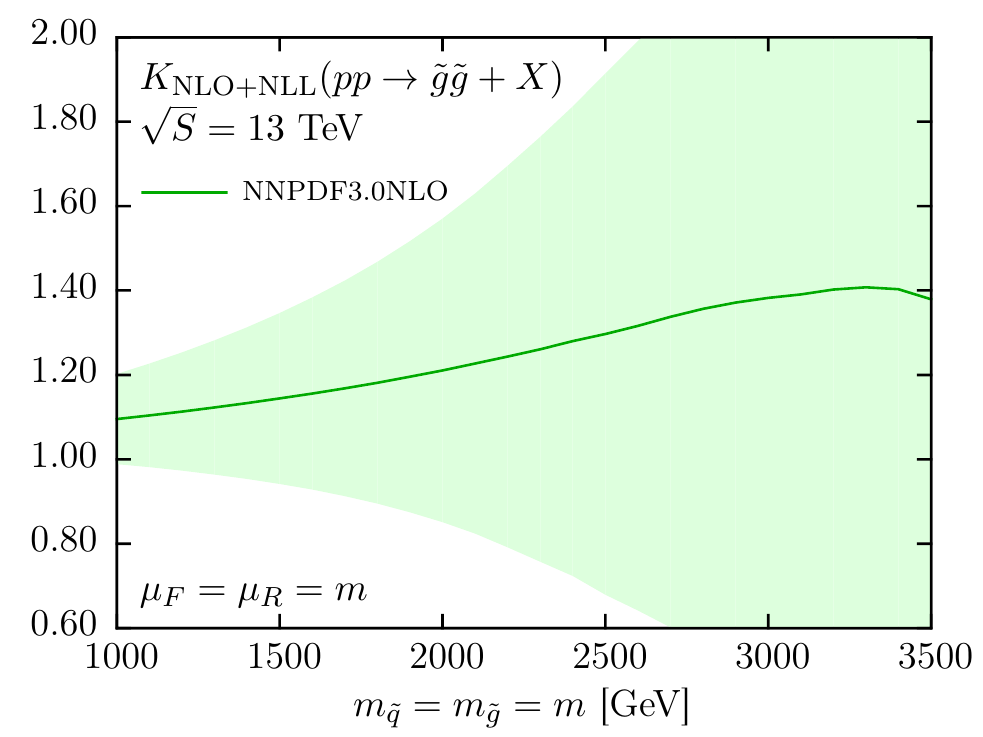}
                \hfill\includegraphics[width=.49\textwidth]{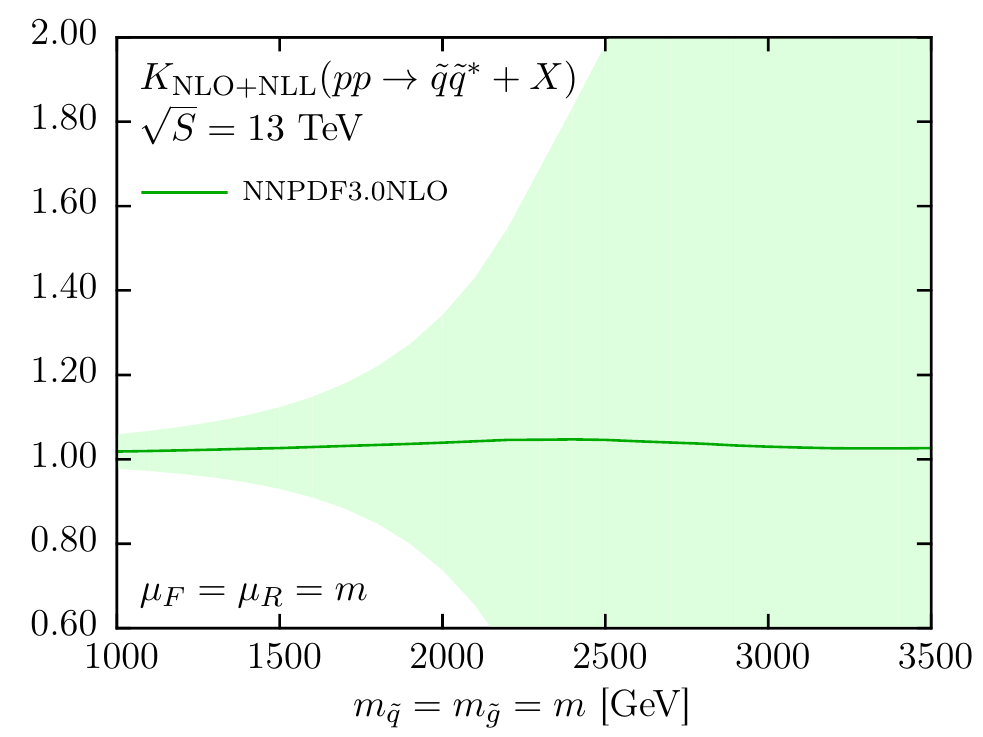}\\
		\includegraphics[width=.49\textwidth]{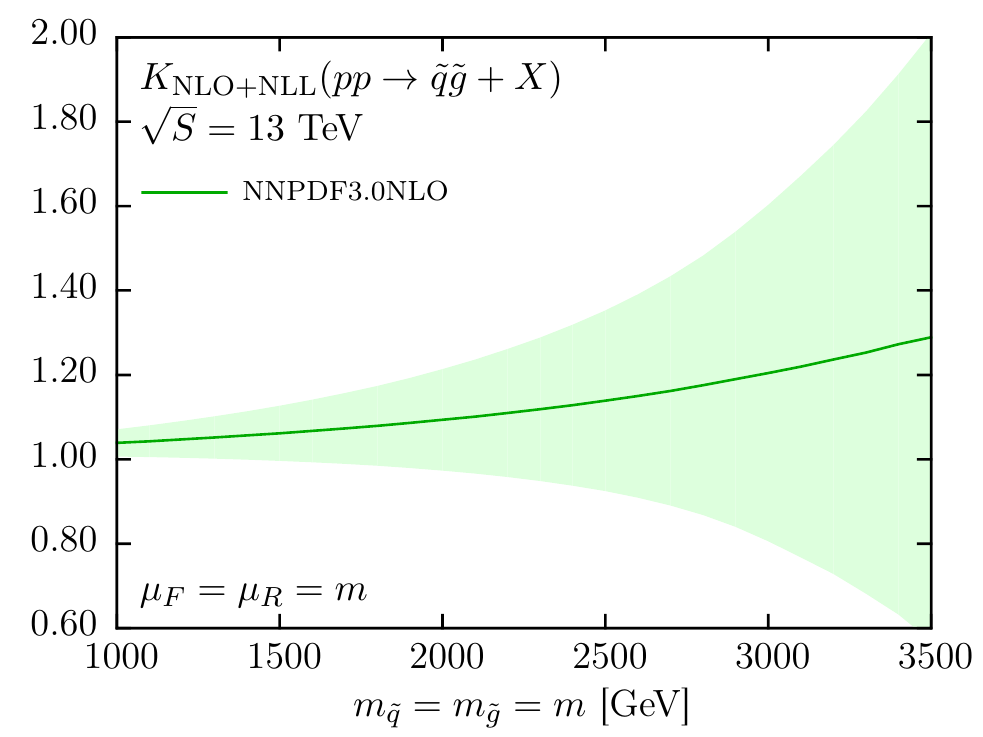}
                \hfill\includegraphics[width=.49\textwidth]{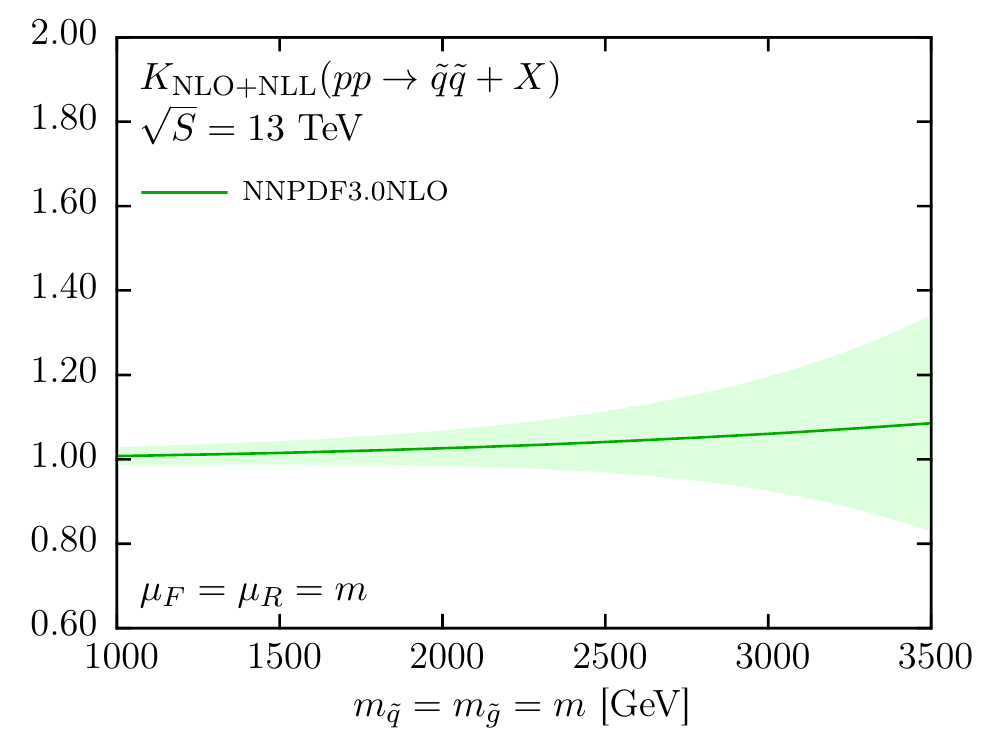}
    \caption{ \small \label{fig:kfact1} The $K$-factor Eq.~(\ref{eq:kfact1}) between the
NLO+NLL and the NLO cross-sections, computed in both cases with the
NNPDF3.0 NLO global set.
We also show the corresponding one-sigma PDF uncertainties.
    }
 	\end{figure}

  In order to illustrate better the size of each of the different effects that enter the calculation, it is
  often useful to represent the cross-sections of Fig.~\ref{fig:xsec1} in terms of a $K$-factor by normalizing to the
  corresponding NLO cross-section (with the same input PDF set), that is
	\be
	K := \frac{\sigma^{\mathrm{NLO+NLL}}\Big|_{\text{NLO global}}}{\sigma^{\mathrm{NLO}}\Big|_{\text{NLO global}}} \, .
  \label{eq:kfact1}
	\ee
  The deviations of this $K$-factor from one indicate the impact of the NLL resummation in the
  partonic cross-sections.
  These $K$-factors are shown in Fig.~\ref{fig:kfact1}, where
  both  the
NLO+NLL and the NLO cross-sections have been computed  with the
NNPDF3.0 global set.
In the computation of Eq.~(\ref{eq:kfact1}), PDF uncertainties are included
in the numerator only.

  These results quantify the effect of the NLL resummation in the partonic cross-sections,
  for the different initial states and as a function of the sparticle mass $m$.
  For instance, for the $\tilde{g}\tilde{g}$ final state, NLL resummation leads
  to an enhancement of the total cross-section that increases from 10\% at
  $m=1$ TeV to 40\% at $m=3$ TeV.
  For the $\tilde{q}\tilde{g}$ case, resummation increases the NLO cross-section by 5\% at
  1 TeV and by 20\% at 3 TeV.
  The effects of threshold resummation are milder for the $\tilde{q}\tilde{q}$
  and $\tilde{q}\tilde{q}^*$ final states,
  where the size of the enhancement is only a few percent.
  Note in all cases the substantial PDF uncertainties at large masses, specially
  for the $\tilde{g}\tilde{g}$ and $\tilde{q}\tilde{q}^*$ final states.

\begin{figure}[t]
	  \centering
          \includegraphics[width=.49\textwidth]{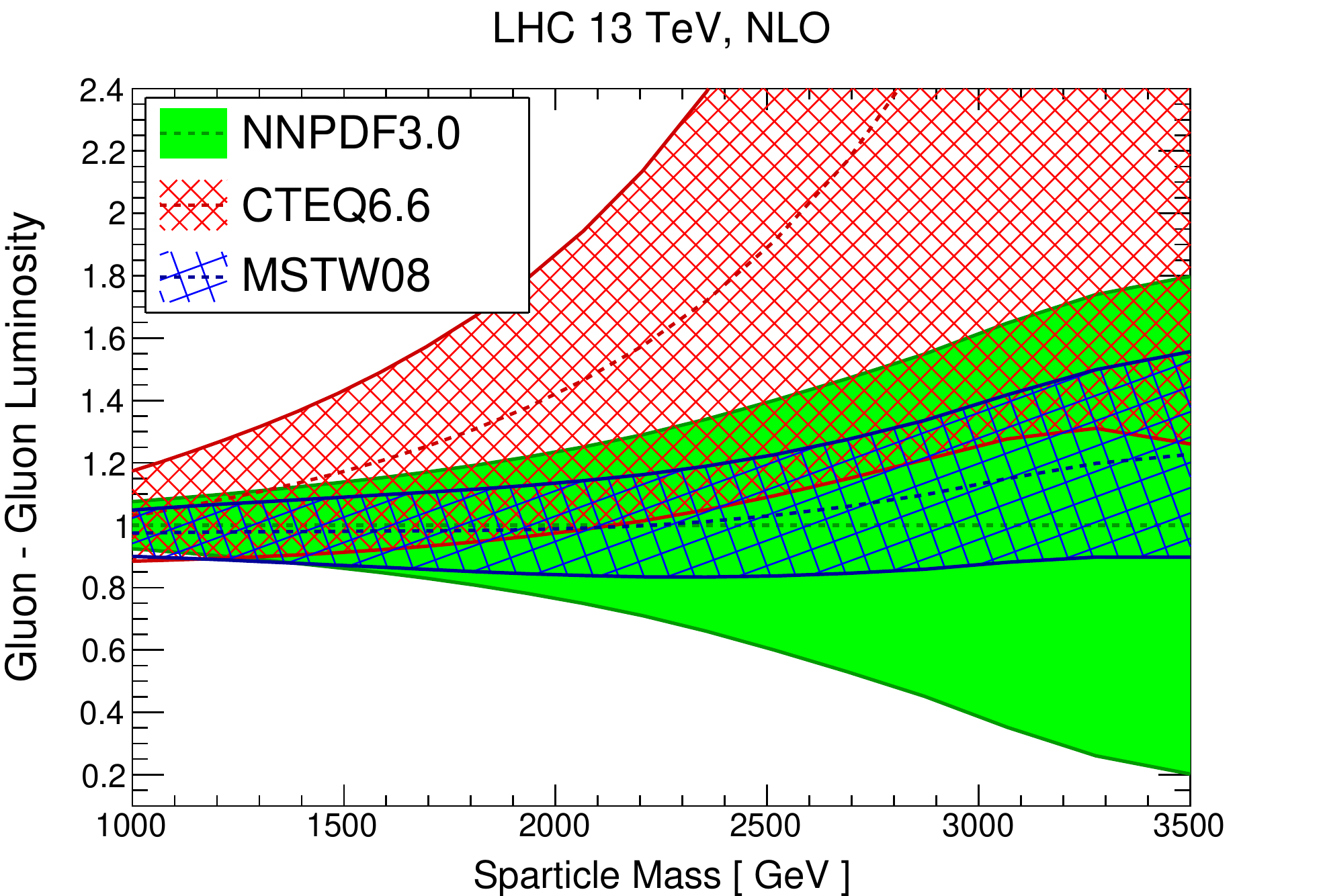}\hfill
                \includegraphics[width=.49\textwidth]{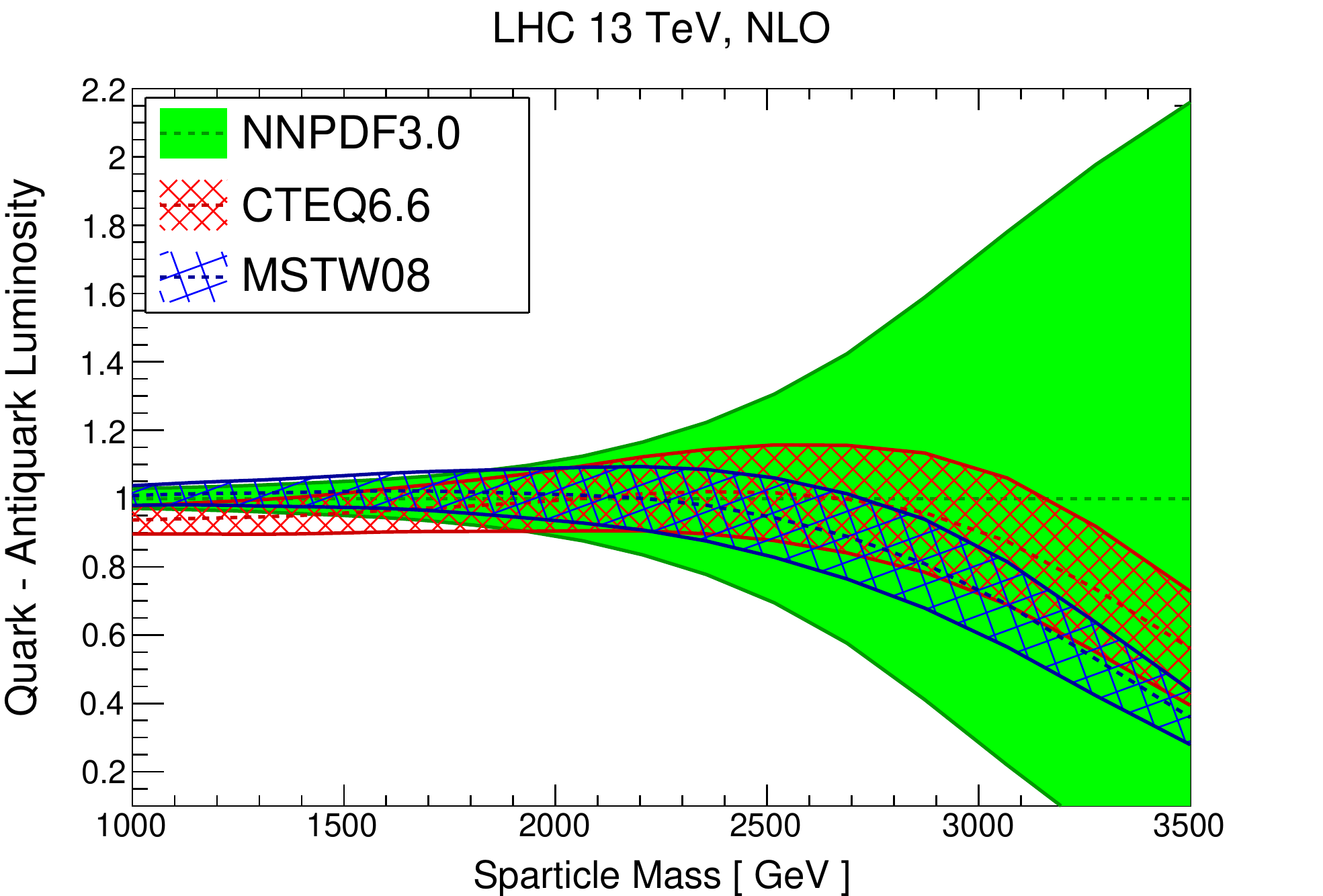}\\
                      \includegraphics[width=.49\textwidth]{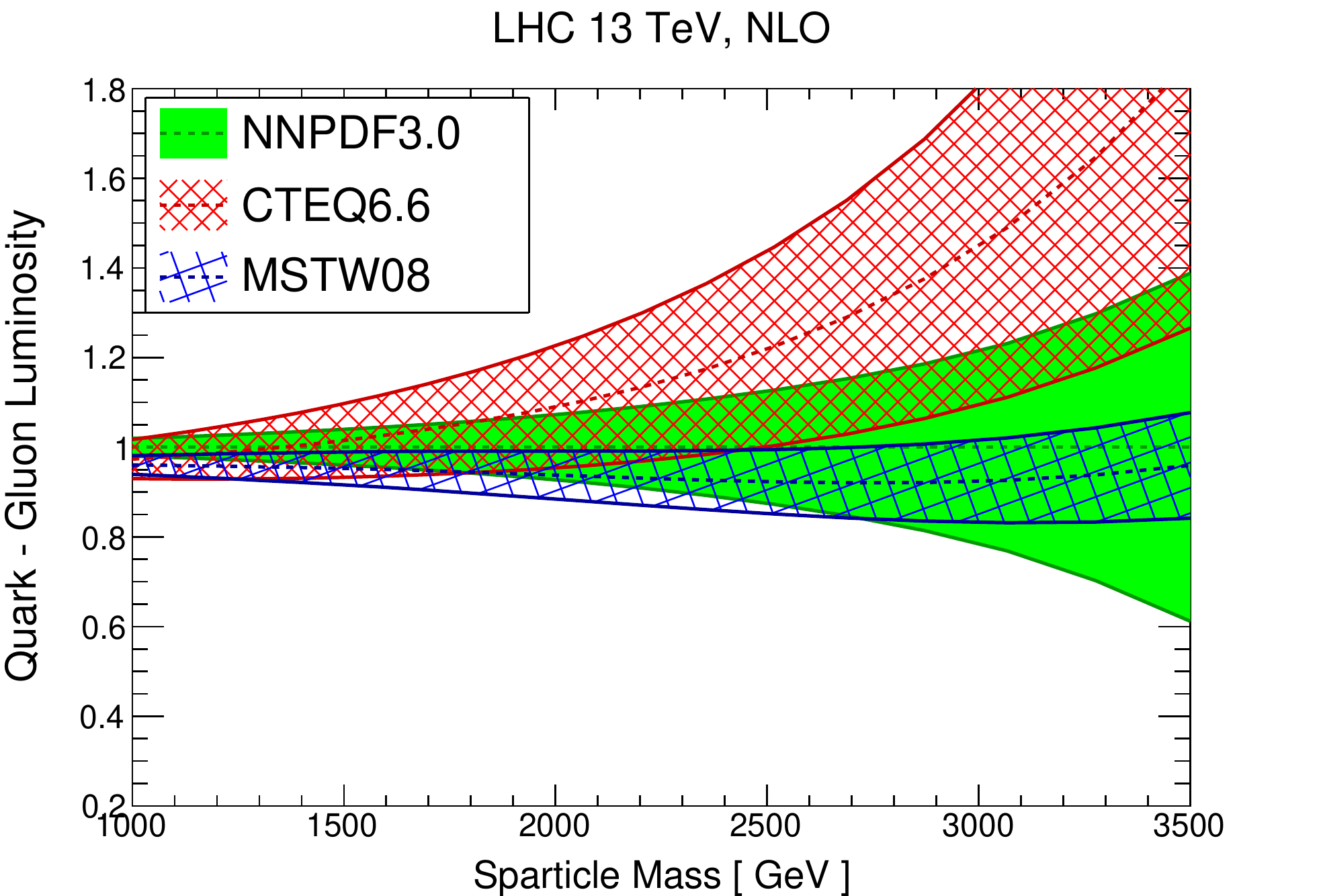}\hfill
		      \includegraphics[width=.49\textwidth]{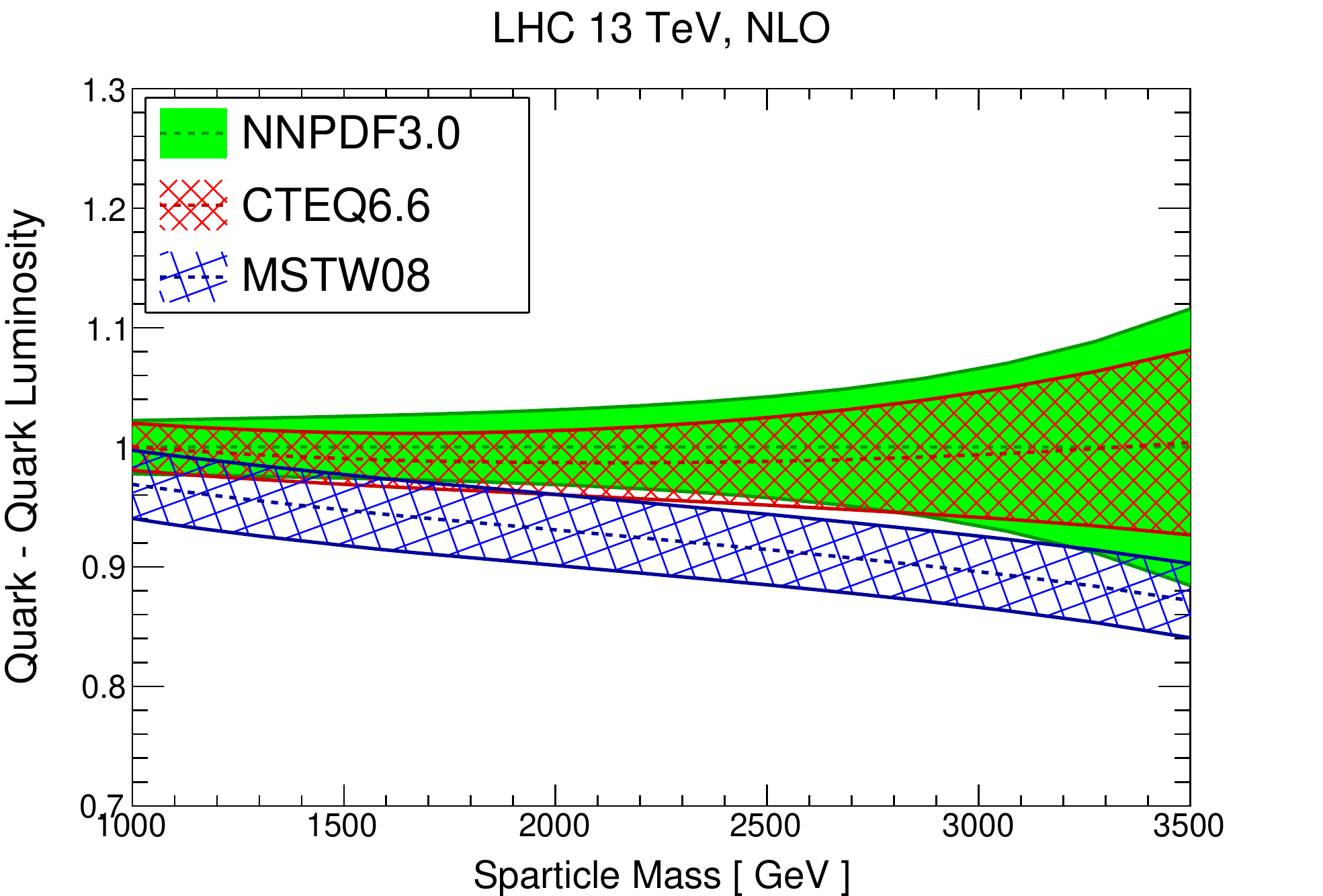} \\
    \caption{\label{fig:pdflumi_old} \small
Comparison of PDF luminosities between the NLO NNPDF3.0, CTEQ6.6
and MSTW08 PDF sets, as a function of the
sparticle mass $m$.
From top to bottom and from left to right we show the gluon-gluon, quark-antiquark, quark-gluon,
and quark-quark PDF luminosities, normalized to the central value of the NNPDF3.0 result.
    }
	\end{figure}

  It is useful to quantify how these updated
  results for the NLO+NLL cross-sections,
  based on NNPDF3.0 NLO, differ from those previously available,
  which were computed using the CTEQ6.6 and MSTW08 NLO PDF sets.
  We study these differences, both for central values and for the
  total PDF uncertainties, first at the level of PDFs and then at the level of 
  of $K$-factors for the resummed cross-section.
  First of all let us compare in Fig.~\ref{fig:pdflumi_old}
  the PDF luminosities between NNPDF3.0, CTEQ6.6
  and MSTW08 NLO -- this comparison is more useful than that of the individual
  PDFs, since differences in the PDF luminosities directly
  translate  into differences in the predicted
  SUSY cross-sections.

  For the gluon-initiated processes, we find good agreement for the central
  values of NNPDF3.0 and MSTW08, with the PDF uncertainties of the latter
  being significantly smaller, especially at high masses.
  On the other hand, the sizes of the
  PDF uncertainties in the $gg$ and $qg$ luminosities are
  comparable
  between NNPDF3.0 and CTEQ6.6, with the latter exhibiting a much harder
  large-$x$ gluon: at $m=2.5$ TeV, the $gg$ luminosity from CTEQ6.6 is a factor
  2 larger than that of NNPDF3.0 (though still consistent within uncertainties).
  For the $qq$ luminosity, there is good agreement between NNPDF3.0 and CTEQ6.6
  in terms of both central values and uncertainties, with MSTW08 being
  systematically smaller by up to 10\%.
  Finally, for the $q\bar{q}$ luminosity, there is reasonable
  agreement in the central value of the three sets, but for
  masses $m\gsim 2$ TeV the PDF uncertainty in NNPDF3.0 is much
  larger than that of the two other PDF sets.
  
   The agreement between NNPDF3.0 and the most recent updates of the CTEQ6.6 and
  MSTW08 sets, namely CT14~\cite{Dulat:2015mca} and MMHT14~\cite{Harland-Lang:2014zoa}, is improved as compared to what is shown
  in Fig.~\ref{fig:pdflumi_old}. In particular now the NNPDF3.0 PDF uncertainty band essentially includes
  the luminosities from CT14 and MMHT14.
  Therefore, the results presented in this work would not change substantially
  if the combination of NNPDF3.0, CT14 and MMHT14 was used instead of only NNPDF3.0
  as we do now.

\begin{figure}[t]
	  \centering
     \includegraphics[width=.49\textwidth]{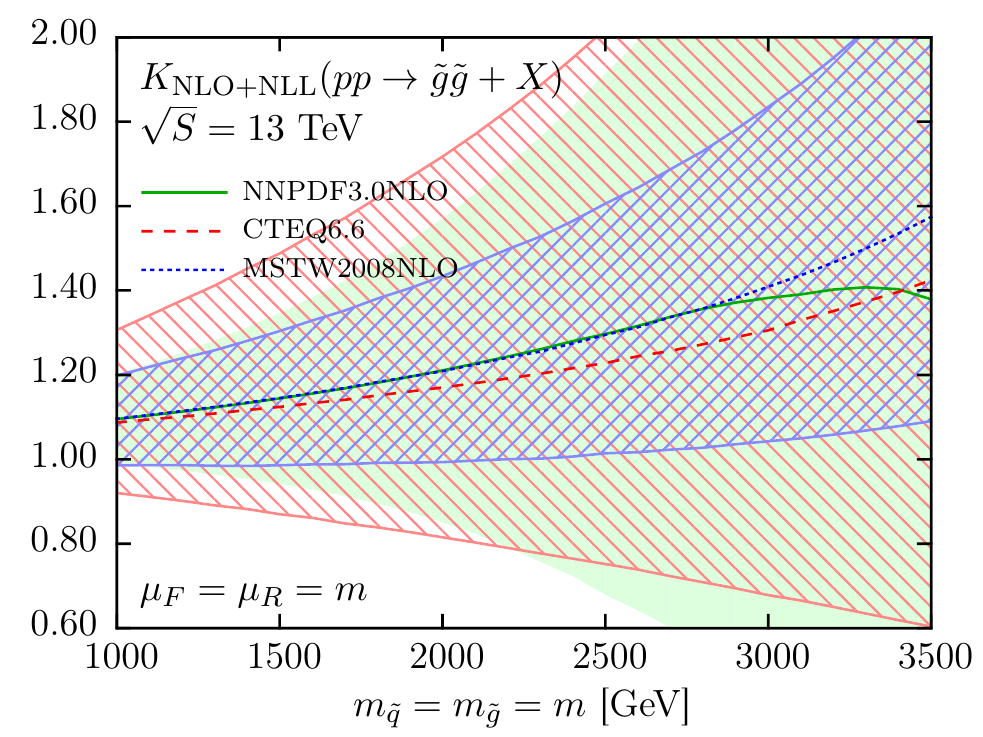}\hfill
     \includegraphics[width=.49\textwidth]{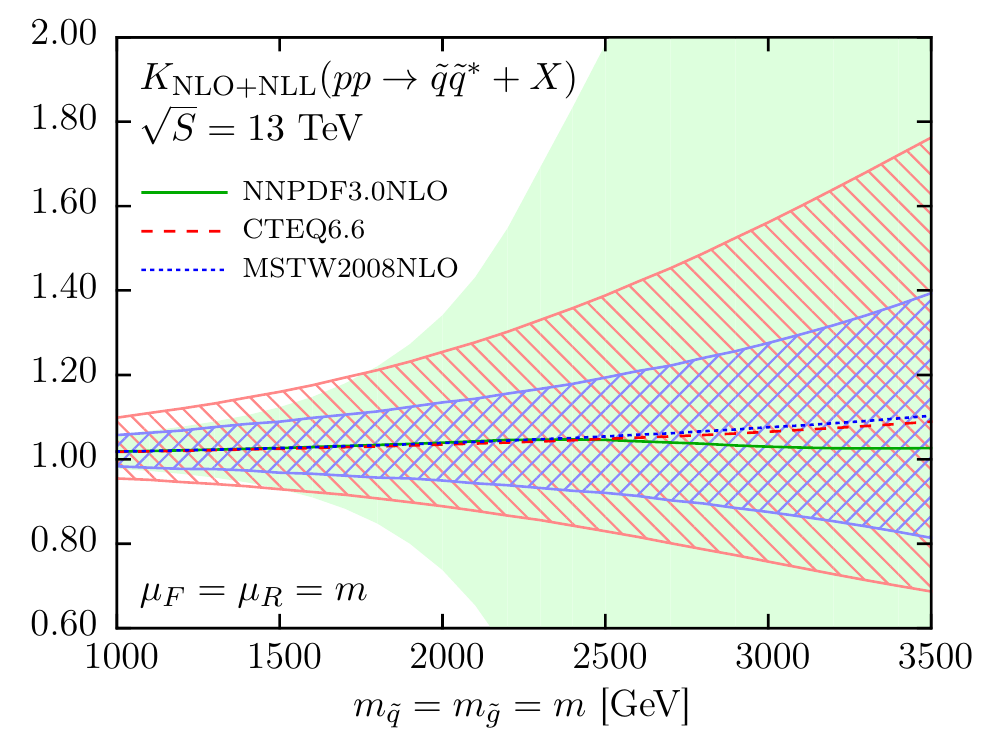}\\
     \includegraphics[width=.49\textwidth]{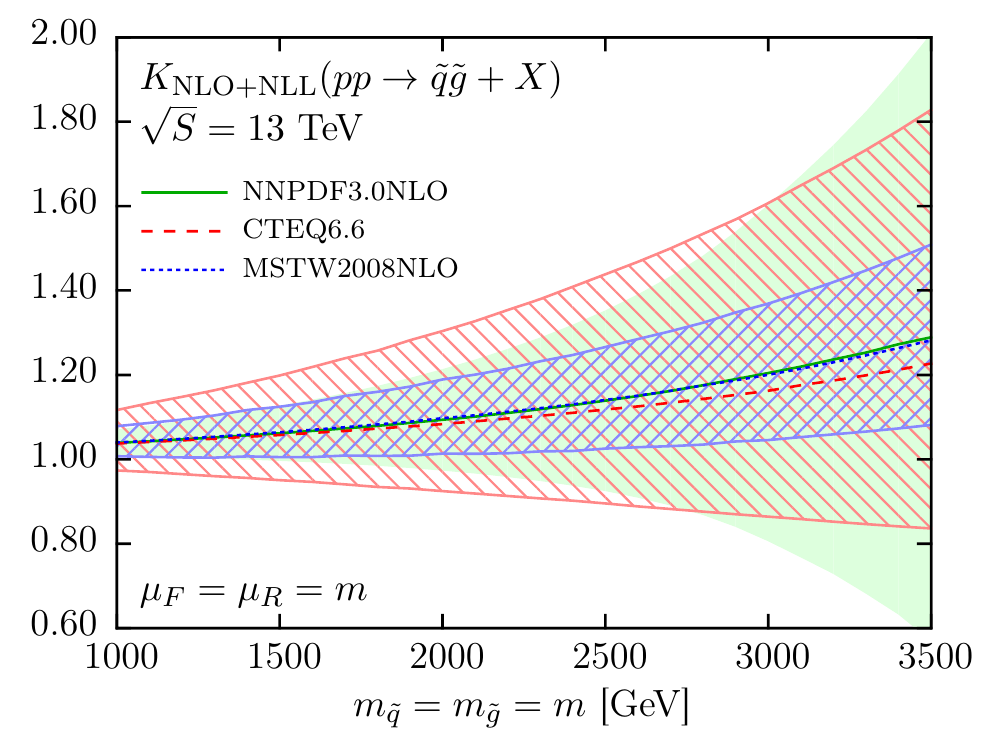}
     \hfill\includegraphics[width=.49\textwidth]{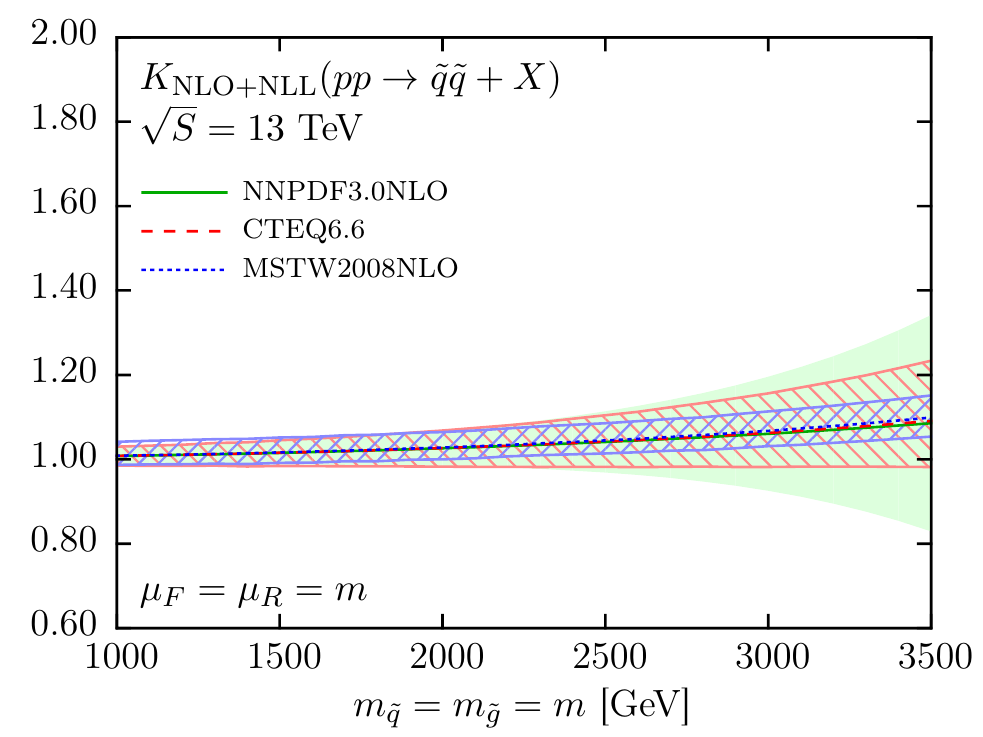}
     \caption{ \small \label{fig:kfact1all}
 Same as Fig.~\ref{fig:kfact1}, now including also the results obtained
 using CTEQ6.6 and MSTW08 NLO, including the corresponding
 PDF uncertainties in each case. 
    }
 	\end{figure}

  %
  Then in Fig.~\ref{fig:kfact1all} we show
  the $K$-factor Eq.~(\ref{eq:kfact1}) between the
  NLO+NLL and the NLO cross-sections, computed with
  NNPDF3.0, CTEQ6.6 and MSTW08 NLO, with the corresponding PDF uncertainties in each case.
  In computing the $K$-factor Eq.~(\ref{eq:kfact1}) we use the central  member both in the numerator and denominator, but the error members (either
  replicas or eigenvectors) only in the numerator, to compute the  PDF uncertainty.
Thus, most of the differences in the central values observed
  in the luminosity comparison of Fig.~\ref{fig:pdflumi_old} will cancel out.
  On the other hand,
  the PDF uncertainties at the $K$-factor level should be consistent with
  those at the PDF luminosity level.

    From Fig.~\ref{fig:kfact1all} we see that that for the
  final states which are predominantly gluon-initiated,
  $\tilde{g}\tilde{g}$ and $\tilde{q}\tilde{g}$, the size of the PDF error bands
  is similar between NNPDF3.0 and CTEQ6.6, being somewhat smaller for
  MSTW08.
  The central values obtained from the three PDF sets are very similar,
  except at large values of the sparticle masses $m$
  where the CTEQ6.6 prediction deviates a bit
  from that of the other two sets.
  The size of the PDF uncertainties observed in
  Fig.~\ref{fig:kfact1all} is consistent with the
   comparison of the $gg$ and $qg$ PDF
   luminosities in Fig.~\ref{fig:pdflumi_old}.
  For quark-initiated processes we find a similar consistency with the
  PDF uncertainties exhibited by the $qq$ and $q\bar{q}$ luminosities.
  In particular, the size of the
  PDF uncertainty band for $\tilde{q}\tilde{q}^*$ production is rather
  larger in NNPDF3.0 as compared to MSTW08 and CTEQ6.6,
  while for $\tilde{q}\tilde{q}$ production the three PDF
  sets give similar results.

\begin{figure}[t]
		\centering
		\includegraphics[width=.49\textwidth]{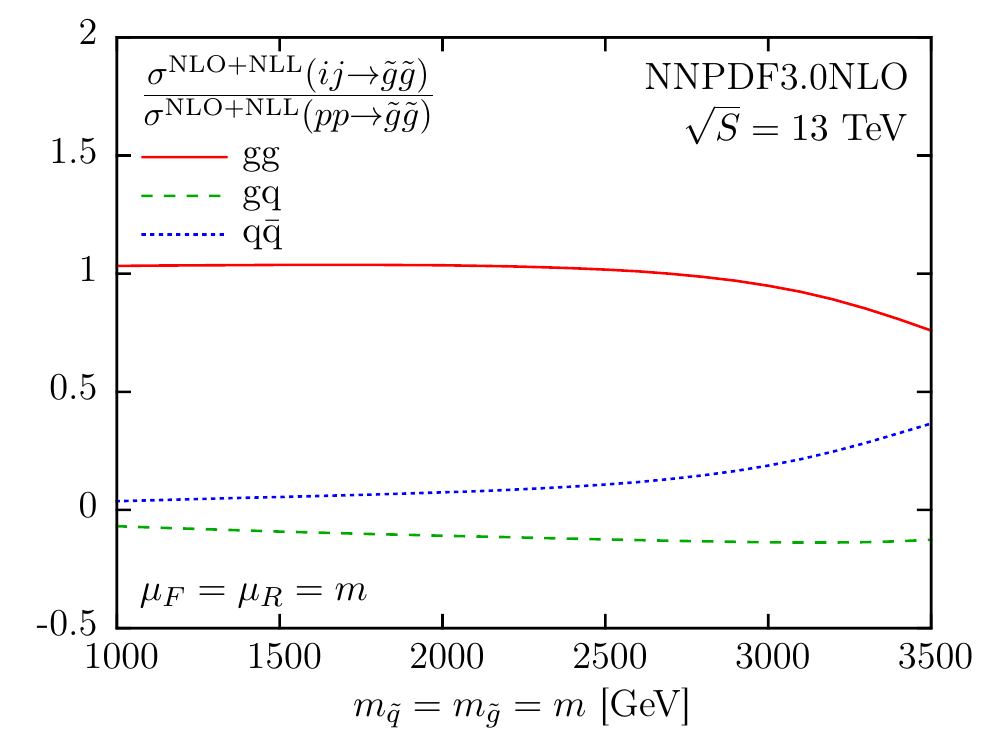}\hfill
    \includegraphics[width=.49\textwidth]{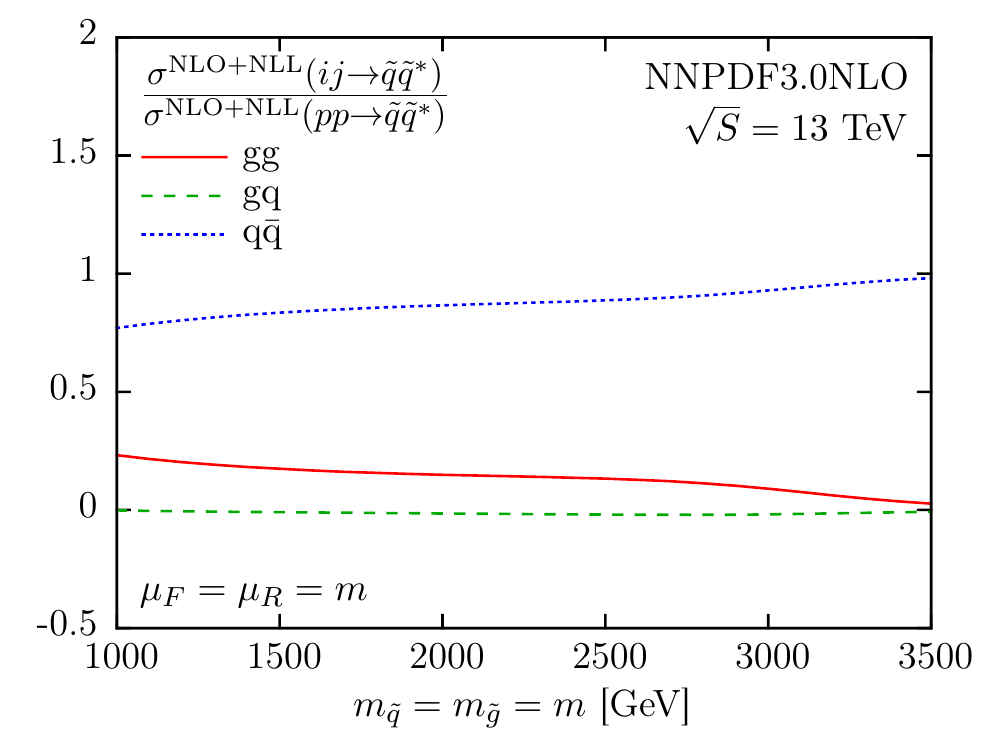}\\
		\includegraphics[width=.49\textwidth]{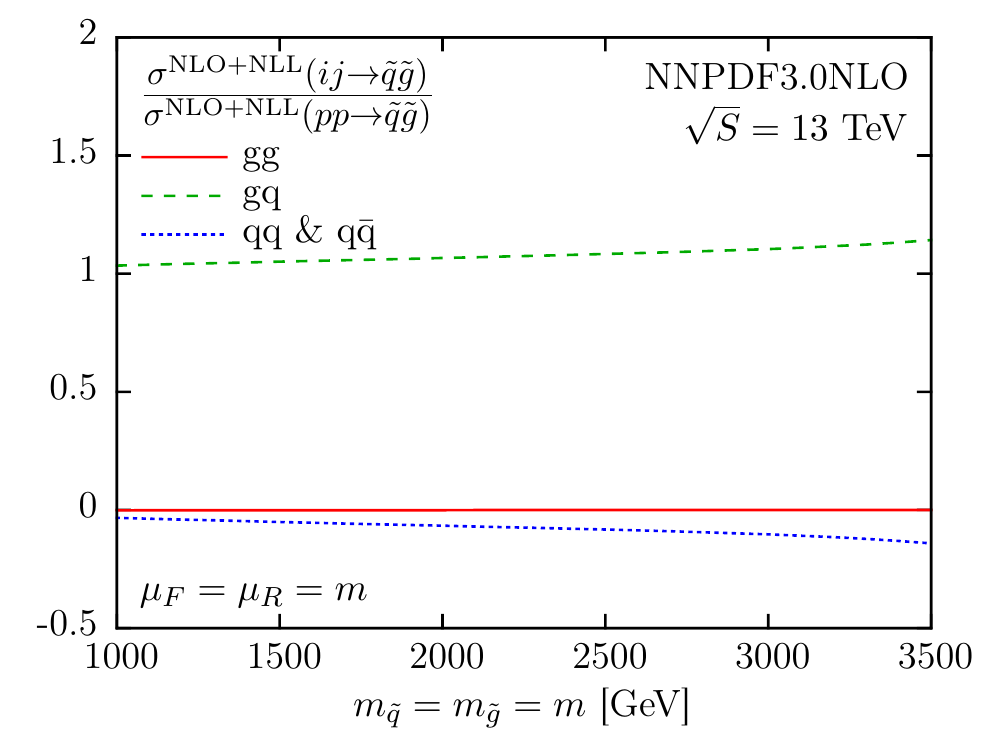} \hfill
    \includegraphics[width=.49\textwidth]{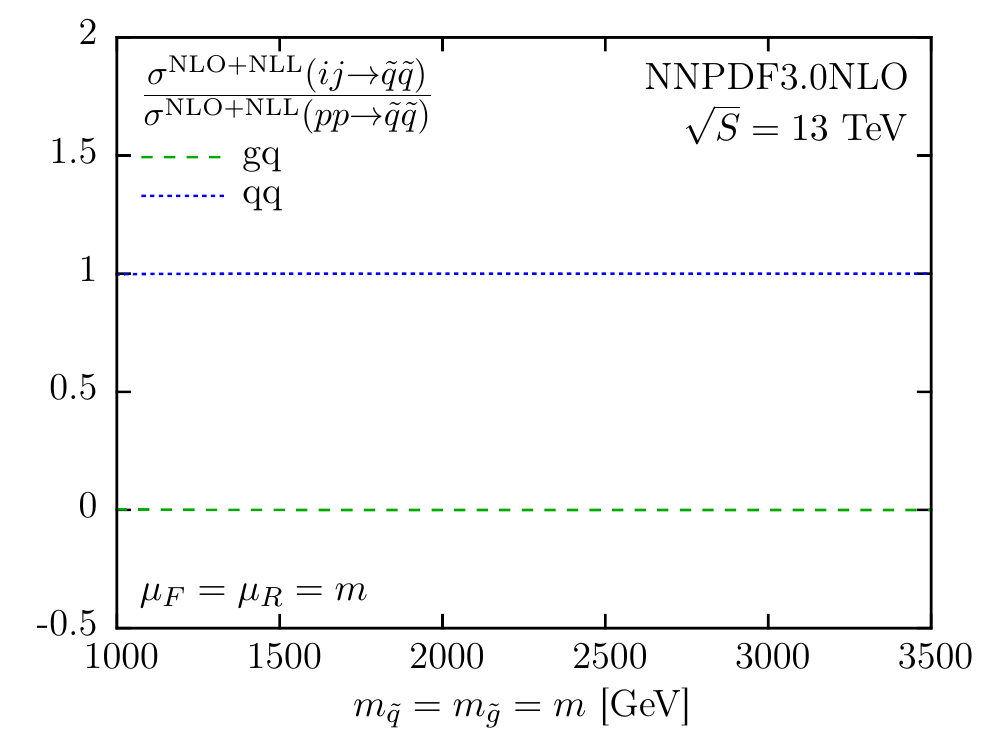}
    \caption{\label{fig:initial_state_splitup} \small Relative contributions from various initial states to the  NLO+NLL squark and gluino cross-sections, Eq.~(\ref{eq:fraction}),
      calculated with NLO NNPDF3.0 global analysis.}
	\end{figure}

  It is interesting to understand the results of Figs.~\ref{fig:xsec1} and~\ref{fig:kfact1}
  in terms of the
  decomposition of the initial state into the different components that contribute to the production of each final state.
  This decomposition is shown in Fig.~\ref{fig:initial_state_splitup}, where
  we show the ratio
  \be
  \label{eq:fraction}
\frac{\sigma^{\rm NLO+NLL}(ij \to kl)}{\sigma^{\rm NLO+NLL}(pp \to kl)} \, ,
\ee
where $ij$ are the different initial-state partonic luminosities, and
$kl$ label each of the final states.

We observe that
the $\tilde{q}\tilde{g}$ and $\tilde{q}\tilde{q}$ final states
are produced entirely from the $qg$ and $qq$ initial states respectively (these
are the only initial states allowed at LO). 
Therefore, predictions for these two final states will follow
closely the behaviour of the corresponding PDF luminosities.
The $\tilde{g}\tilde{g}$ final state receives contributions both from the $q\bar{q}$ and $gg$ initial states,
but it
is the latter that dominates and  determines the behaviour of the PDF uncertainty band
through the $gg$ luminosity, cf. Fig.~\ref{fig:pdflumi_old}.
Similarly, the size and the shape of the PDF uncertainties for $\tilde{q}\tilde{q}^*$ production reflects
the properties of the $q\bar{q}$ luminosity as shown in Fig.~\ref{fig:pdflumi_old}.

  To summarize, in this section we have presented updated
  results for squark
  and gluino pair production cross-sections at NLO+NLL
  accuracy~\cite{Beenakker:2009ha}.
  So far, available predictions~\cite{Beenakker:2009ha,Beenakker:2011fu, Kramer:2012bx, Borschensky:2014cia} were obtained using CTEQ6.6 and MSTW08, while here the recent NNPDF3.0 NLO global fit is used.
  The combination of an unbiased PDF parametrization with the constraints from all available hard-scattering data, including a number of LHC measurements, make NNPDF3.0 specially suitable
  to provide a robust estimate of PDF uncertainties for the production of massive
  particles in the TeV region.

  The updated NLO+NLL SUSY cross-sections using the NNPDF3.0 NLO
  global analysis are available
  from the webpage of the {\tt NLL-fast} collaboration~\cite{NNLfastWebpage}
  in the format of fast interpolation grids.
  These grids provide,
  for any value of the squark
  and gluino masses in the range relevant for
  LHC applications, the NLO and NLO+NLL cross-sections
  together with the overall theory uncertainty, separated in the scale, PDF
  uncertainties and $\alpha_s$ uncertainties.
  In the latter case, we assume
$\delta \alpha_s=0.0012$ at the 68\% confidence level, and
  follow the PDF4LHC prescription~\cite{Botje:2011sn}
  for the combination of PDFs and $\alpha_s$ uncertainties for
  Monte Carlo PDF sets~\cite{Demartin:2010er}.

%% file: sec-resummation.tex
\section{Impact of threshold-improved PDFs on the NLO+NLL cross-sections}
\label{sec-resummation}

Now we discuss how the NLO+NLL results of the previous section, obtained
with the NNPDF3.0 NLO global fit as input, are modified when
threshold-improved NLO+NLL PDFs are used instead.
The main subtlety here arises from the fact that, as explained in the introduction,
the NNPDF3.0 NLO+NLL resummed sets are based on a smaller dataset than
their NNPDF3.0 NLO counterparts.
Therefore, it is required to devise a prescription for
combining results
of the fixed order global fit, which is the most precise in terms of
experimental constraints, and the resummed fit, which is based on a more accurate
theory but has larger PDF uncertainties.
It is also important to emphasize that
to  assess  consistently the impact of NLO+NLL threshold
resummation as compared to a NLO fixed-order fit, one should always
use PDF sets based on exactly the same input dataset.
In the following, using the convention
of~\cite{Bonvini:2015ira}, we will denote by ``DIS+DY+top'' the NNPDF3.0
fixed-order and resummed fits based on this reduced dataset.
%

	\begin{figure}[t]
		\centering
		\includegraphics[width=.49\textwidth]{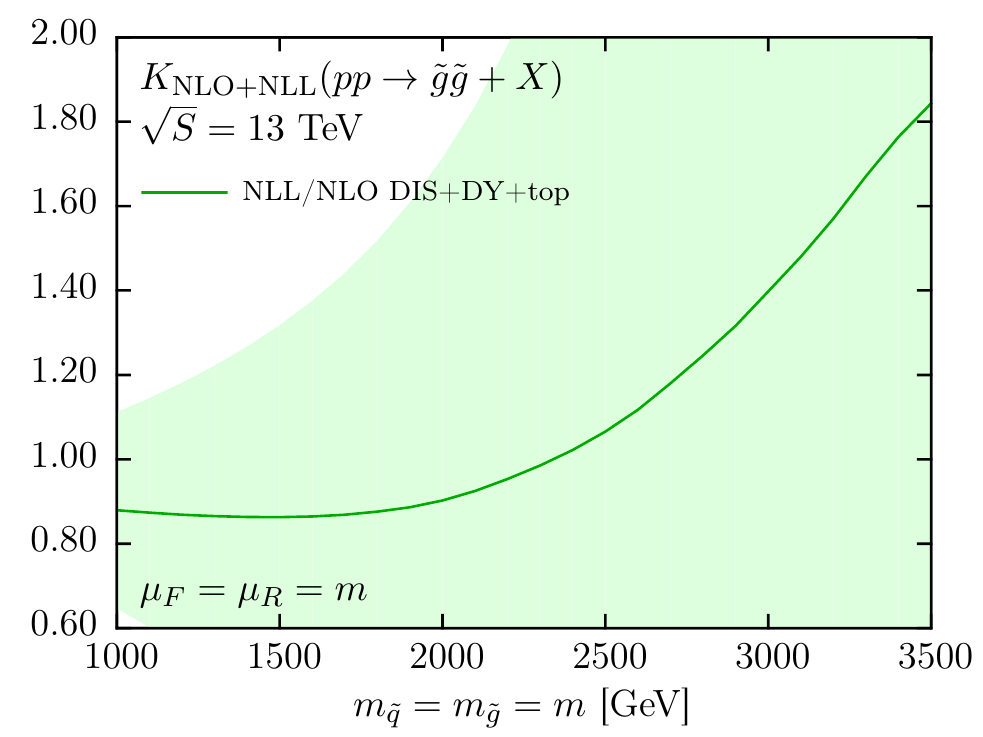}\hfill
                \includegraphics[width=.49\textwidth]{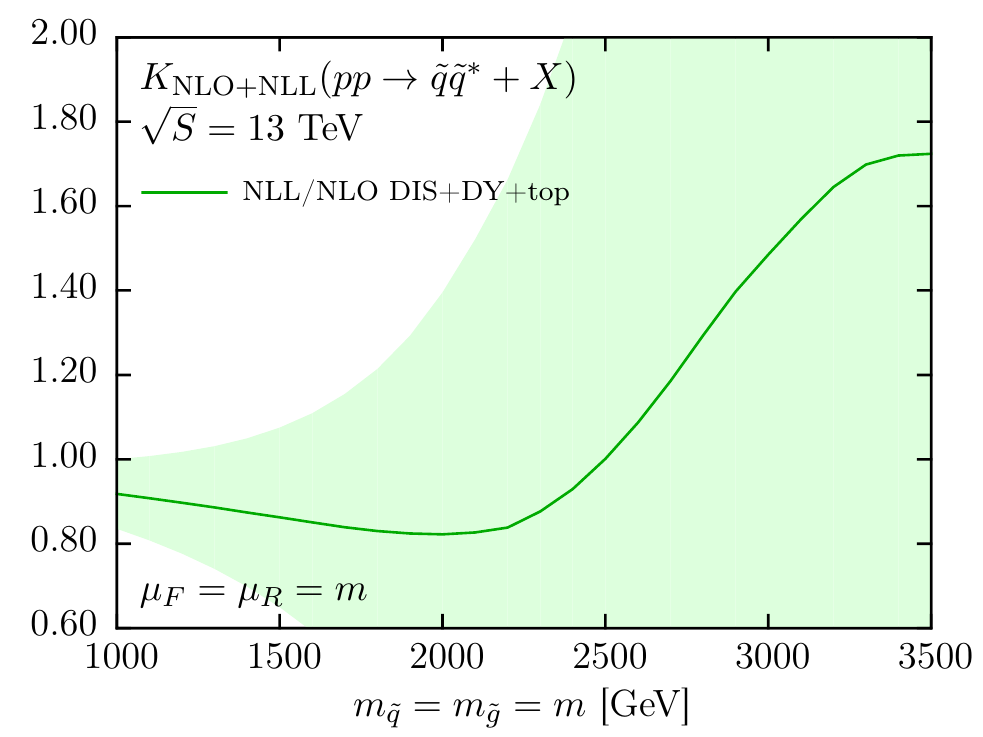}\\
		\includegraphics[width=.49\textwidth]{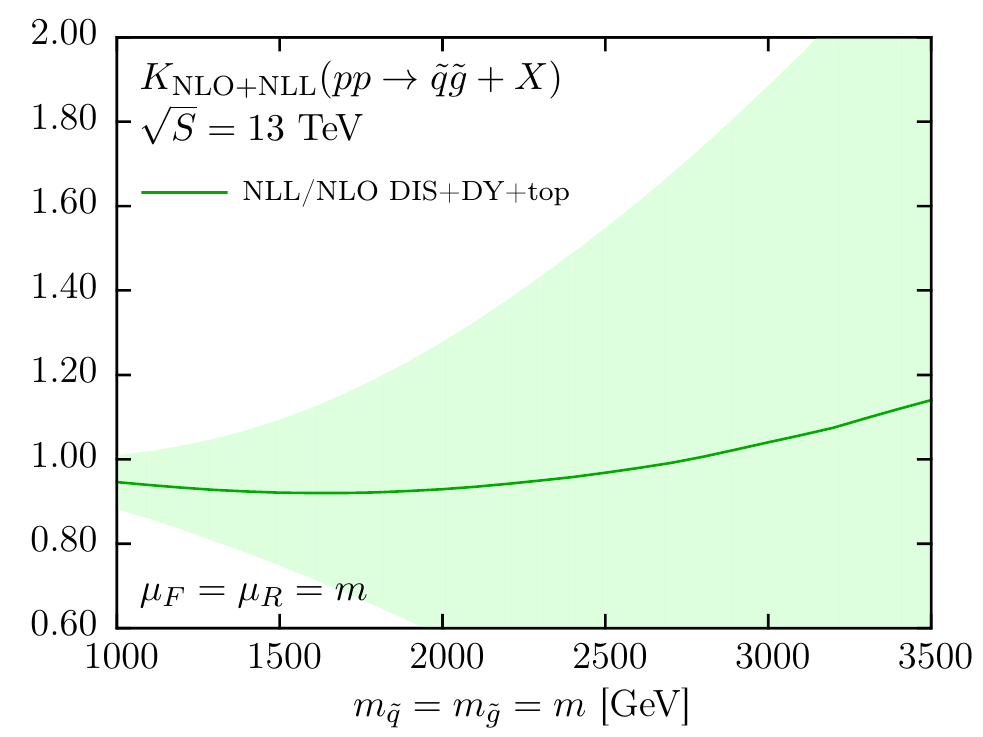}\hfill
                \includegraphics[width=.49\textwidth]{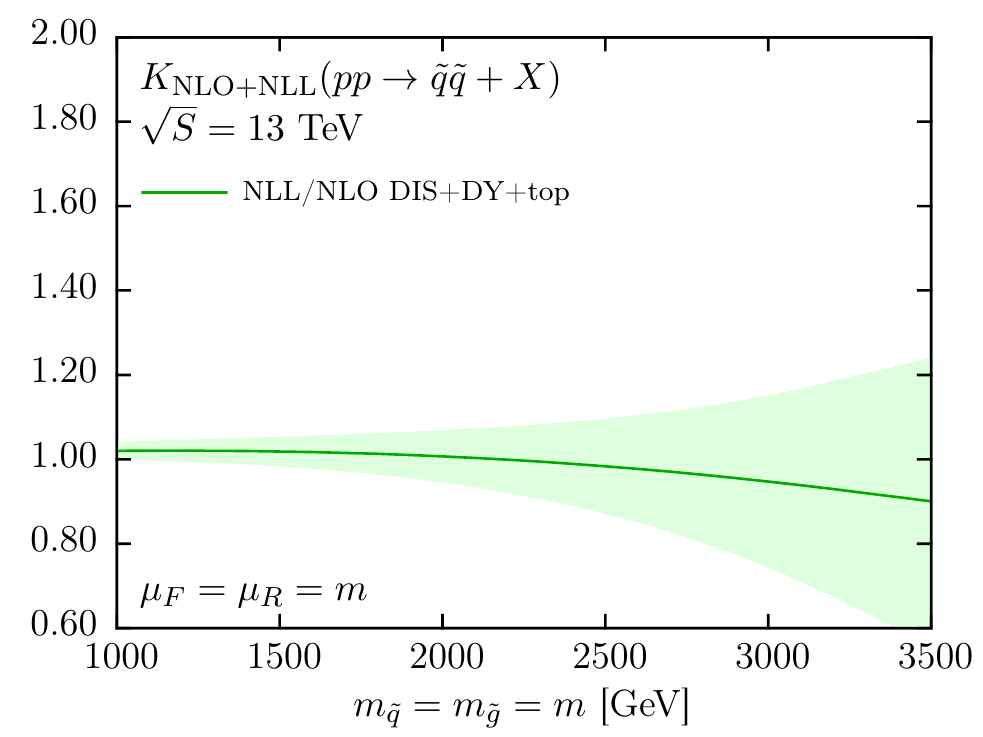}
                \caption{\label{fig:kfact2} \small
Same as Fig.~\ref{fig:kfact1}, but now for the $K$-factor defined in Eq.~(\ref{eq:kfact2}).}
	\end{figure}

Now we quantify, using the same settings as in the previous section,
the impact of threshold resummation on the supersymmetric
particle pair production cross-section, with resummation included for
partonic cross-sections and of parton distributions.
This can be achieved by defining a new $K$-factor as follows:
\be
\label{eq:kfact2}
K := \frac{\sigma^{\mathrm{NLO+NLL}}\Big|_{\text{NLL DIS+DY+top}}}{\sigma^{\mathrm{NLO}}\Big|_{\text{NLO DIS+DY+top}}} \, ,
\ee
where we note the two differences as compared to Eq.~(\ref{eq:kfact1}):
\begin{itemize}
\item we use as input the fixed-order and resummed NNPDF3.0 DIS+DY+top
  fits, rather than the global fits,
\item the perturbative order of the PDFs matches  that
  of the partonic cross-sections: resummed in the numerator, fixed-order
  in the denominator.
\end{itemize}

The results for the $K$-factors defined in Eq.~(\ref{eq:kfact2}) are shown in
Fig.~\ref{fig:kfact2}.
As compared to the corresponding results of  Fig.~\ref{fig:kfact1}, obtained
using only fixed-order PDFs, there are some important differences.
First of all, when a fixed-order PDF is used in the resummed calculation, Fig.~\ref{fig:kfact1},
we found that
the effect of threshold resummation was almost always to increase the total cross-section monotonically
as a function of the sparticle mass.
The only exception was for very high masses for $\tilde{g}\tilde{g}$ and $\tilde{q}\tilde{q}^*$, where PDF uncertainties are very large and the central value of the prediction is affected
by correspondingly
large fluctuations.
On the other hand, when threshold-improved
PDFs are used together with resummed cross-sections, Fig.~\ref{fig:kfact2},
the results are qualitatively different.

For sparticle masses below 2 TeV we  now find a $K$-factor smaller than unity,
except for the $\tilde{q}\tilde{q}$ final state.
The absolute size of the $K$-factors for $m\le 2$ TeV
is around 0.9 for $\tilde{g}\tilde{g}$ and 
$\tilde{q}\tilde{g}$, between 0.8 and 0.9 for $\tilde{q}\tilde{q}^*$, and a few percent
above unity
for $\tilde{q}\tilde{q}$.
This behaviour changes at larger sparticle masses, where the $K$-factors can become much larger,
specially for the  $\tilde{g}\tilde{g}$ and $\tilde{q}\tilde{q}^*$ final states.
The origin for this turnover
is the impact of the resummation in the partonic cross-sections,
which becomes dominant and compensates the suppression due to the resummation on the PDFs.
In any case, PDF uncertainties are very large in this region, above 2 TeV.

	\begin{figure}[t]
		\centering
		\includegraphics[width=.49\textwidth]{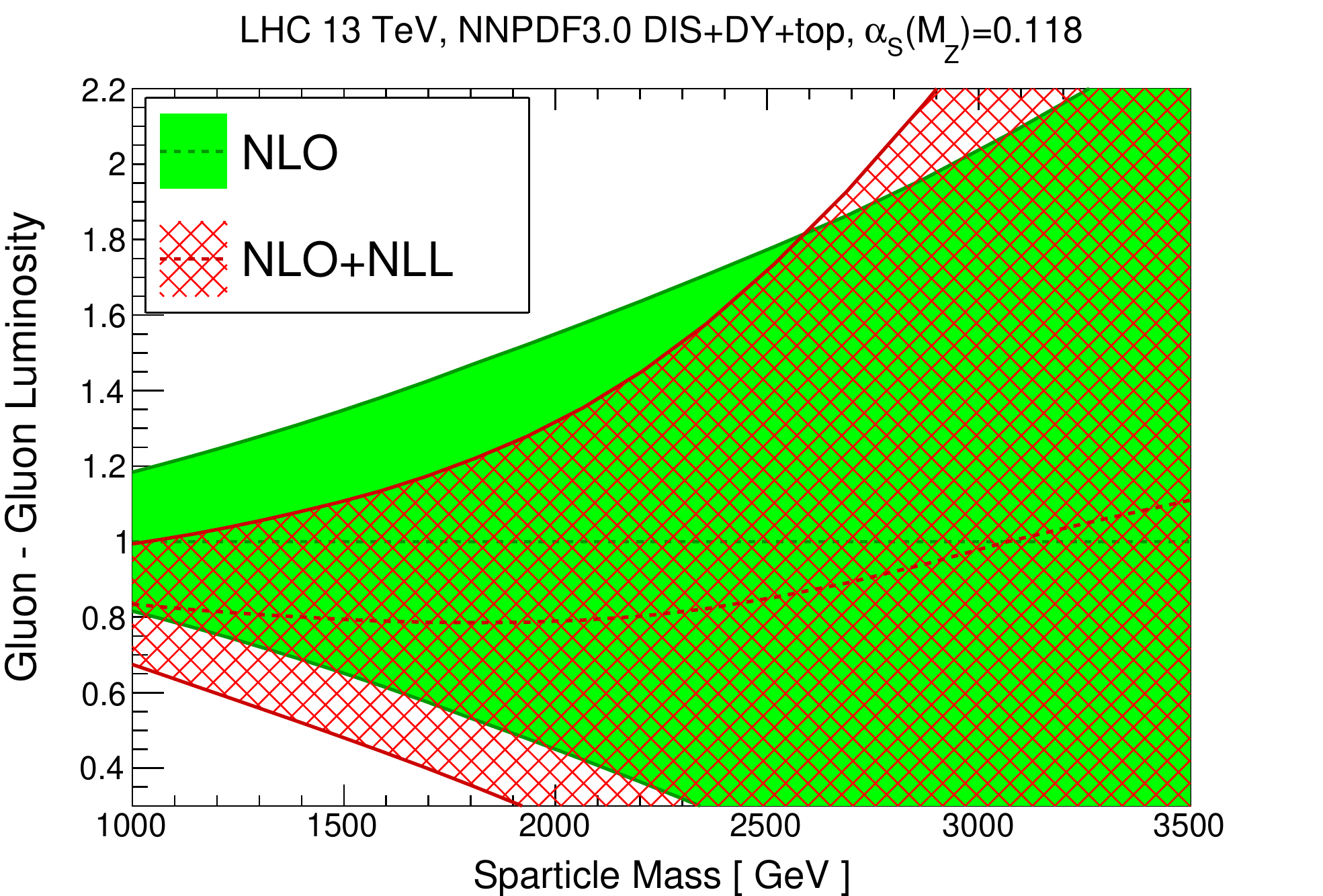}\hfill
                \includegraphics[width=.49\textwidth]{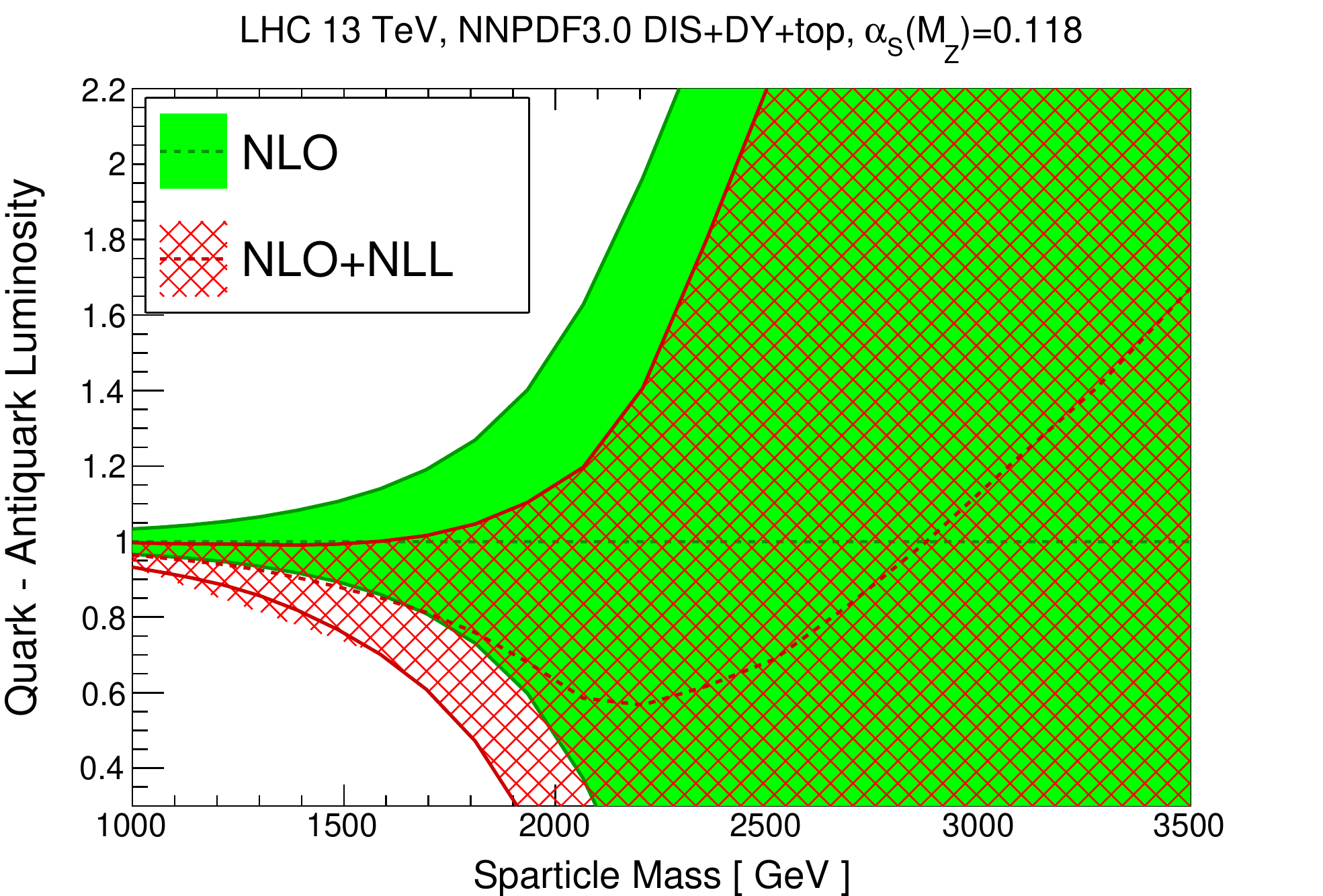}\\
                      \includegraphics[width=.49\textwidth]{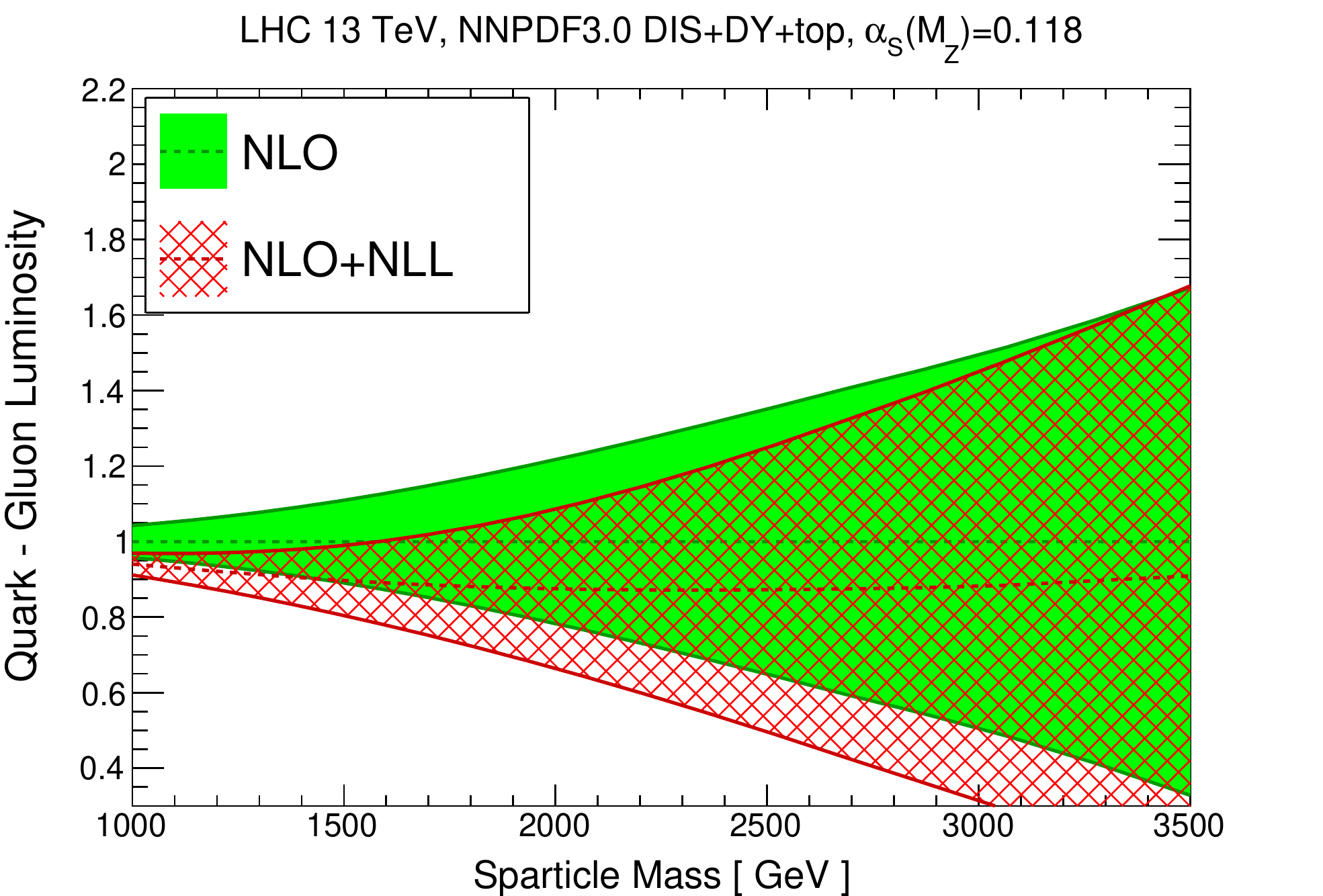}\hfill
		\includegraphics[width=.49\textwidth]{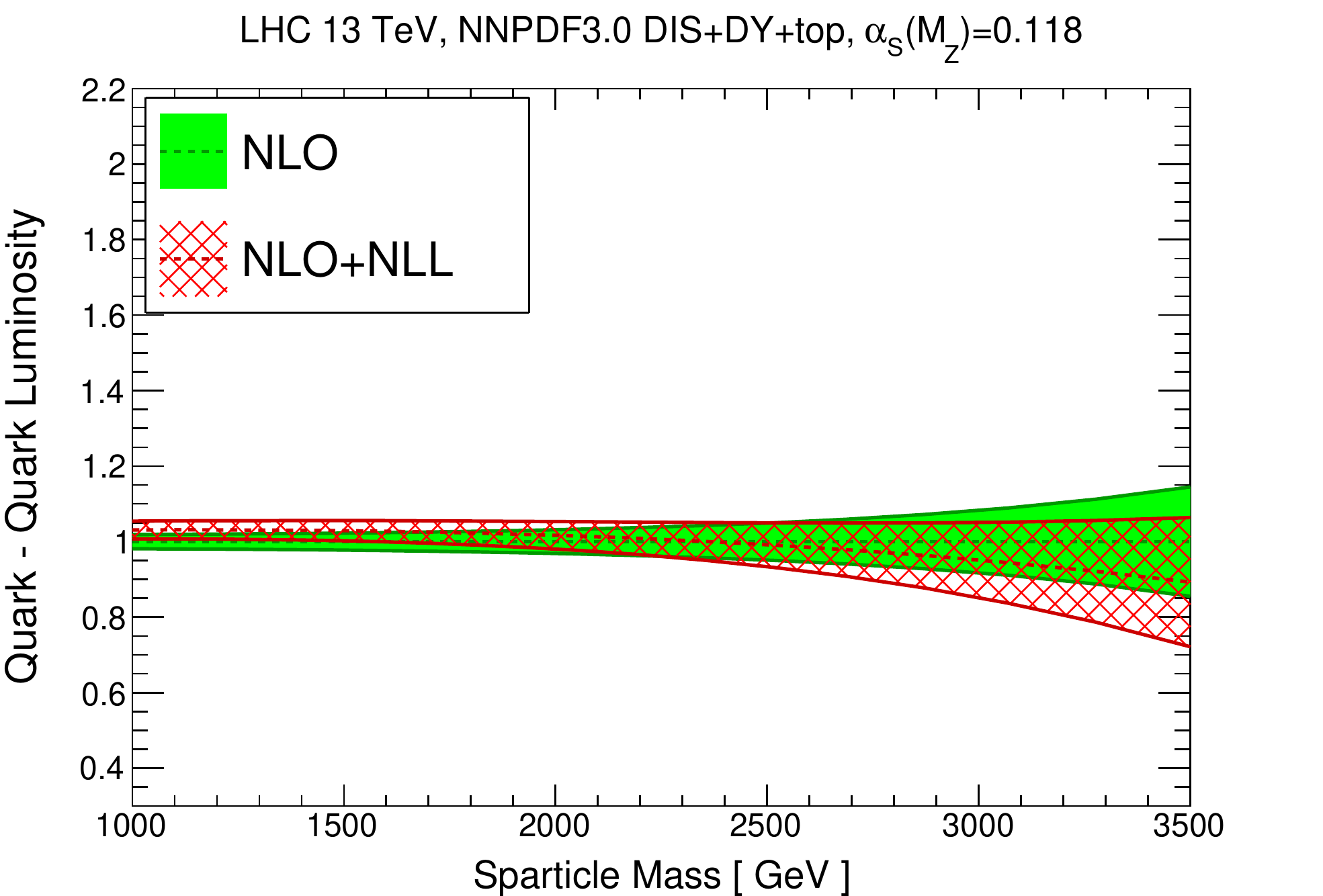} \\
                    \caption{\label{fig:pdflumi} \small
                  Comparison of PDF luminosities between the fixed-order NLO and resummed NLO+NLL NNPDF3.0 DIS+DY+top
            fits, as a function of the produced sparticle mass, in the same range
                  as those of the cross-sections in Fig.~\ref{fig:kfact2}.
                  From top to bottom and from left to right we show the gluon-gluon, quark-antiquark, quark-gluon and quark-quark PDF luminosities, normalized to the central value
                  of the NLO fit.
                }
	\end{figure}

In order to understand these results, it is useful
to compare the DIS+DY+top NLO fixed-order
and NLO+NLL resummed fits in 
terms of partonic luminosities, using exactly the same range of
invariant masses as that shown
in Fig.~\ref{fig:kfact2}.
The results are shown in Fig.~\ref{fig:pdflumi}.
Taking into account the initial PDF combinations that contribute to each final state,
it is possible to quantitatively understand various interesting features.
For example, we see that the $gg$ luminosity is suppressed due to resummation by about 20\%
for masses $m$ between 1 and 2.5 TeV, and then it becomes essentially the same as the fixed
order counterpart within the very large PDF errors.
Correspondingly, towards the lower end of the considered sparticle
mass spectrum, the $K$-factor for $\tilde{g}\tilde{g}$ cross-section
in Fig.~\ref{fig:kfact2} is smaller than in Fig.~\ref{fig:kfact1} by around
$\sim 15\%$, and then
it starts to grow as $m$ is increased when the resummation in the partonic
cross-sections starts to dominate.

Therefore, we conclude that the effect of using
threshold-improved
PDFs in a resummed calculation
cannot
be neglected, and is important to take into account, since it modifies
both the qualitative and quantitative behaviour of the high-mass
sparticle pair production cross-sections at the NLO+NLL accuracy.
        However, the $K$-factors of Fig.~\ref{fig:kfact2} cannot be taken as our
        best results, since they are affected by much larger PDF uncertainties as
        compared to the global
        NNPDF3.0 fit results, cf. Figs~\ref{fig:pdflumi_old} and~\ref{fig:pdflumi}.
        Therefore, we need a prescription to include the effect
        on the SUSY cross-sections of the threshold
        resummation in the PDFs, while keeping all the experimental information on large-$x$ PDFs
        available in the NNPDF3.0 global fit.

        In this respect, we have explored a number of prescriptions to combine the global (fixed-order)
        and DIS+DY+top (resummed) results for the sparticle cross-sections.
        Here we will show results for two possible prescriptions.
        The first is defined as follows:
        \be
        \label{eq:prescription1}
       K := \frac{   \sigma^{\mathrm{NLO+NLL}}\Big|_{\text{NLO global}} }{ \sigma^{\mathrm{NLO}}\Big|_{\text{NLO global}}  }
       \times
       \frac{\sigma^{\mathrm{NLO+NLL}}\Big|_{\text{NLL DIS+DY+top}} }{ \sigma^{\mathrm{NLO+NLL}}\Big|_{\text{NLO DIS+DY+top}}} \, .
       \ee
       It amounts to an overall
       rescaling of the $K$-factor Eq.~(\ref{eq:kfact1}), the result
       obtained using the global NLO set as input, by a factor that accounts
       for the differences in the
       NLO+NLL calculation when using either NLO+NLL or NLO PDFs as input.
       Note that in the limit in which the dataset of the DIS+DY+top fit would become
       identical to that of the global
       (when all observables in the global fit can be simultaneously resummed), this definition automatically reduces
       to the result which would be obtained using a NLO+NLL global fit in the numerator and a NLO global fit in the
       denominator, {\it i.e.},
        \be
       K := \frac{   \sigma^{\mathrm{NLO+NLL}}\Big|_{\text{NLL global}} }{ \sigma^{\mathrm{NLO}}\Big|_{\text{NLO global}}  } \, .
       \ee
       Note that in Eq.~(\ref{eq:prescription1}) PDF uncertainties are not
       included; we are only interested in quantifying the shift
       in the central value of the NLO+NLL cross-sections
       when resummed (rather than fixed-order) PDFs are used
       as input to the calculation.
   
       An improvement with respect to the prescription of Eq.~(\ref{eq:prescription1}) can be achieved by rescaling
       each initial state separately, that is, the overall $K$-factor is now defined as
       \be
       \label{eq:prescription2}
		K = \frac{\sigma^{\mathrm{NLO+NLL}}_{qq}\Big|_{\text{rescaled}}+\sigma^{\mathrm{NLO+NLL}}_{qg}\Big|_{\text{rescaled}}+\sigma^{\mathrm{NLO+NLL}}_{gg}\Big|_{\text{rescaled}}}{\sigma^{\mathrm{NLO}}\Big|_{\text{NLO global}}}
	\ee
	where we have defined the {\it rescaled} cross-sections  for the various initial states
        as follows:
	\be
		\sigma_{ij}^{\mathrm{NLO+NLL}}\Big|_{\text{rescaled}} = \sigma_{ij}^{\mathrm{NLO+NLL}}\Big|_{\text{NLO global}}\times\frac{\sigma_{ij}^{\mathrm{NLO+NLL}}\Big|_{\text{NLL DIS+DY+top}}}{\sigma_{ij}^{\mathrm{NLO+NLL}}\Big|_{\text{NLO DIS+DY+top}}} \, .
	        \ee
                The motivation for this
                second
                prescription Eq.~(\ref{eq:prescription2}) is that it might be more accurate
                to include the effect of resummation in the PDFs separately in each of the individual
                partonic channels.
                As we will now show, from the practical point of view the two prescriptions
                Eq.~(\ref{eq:prescription1}) and  Eq.~(\ref{eq:prescription2}) yield very similar
                numerical results, which indicates that our proposed strategy
                is robust.

\begin{figure}[t]
		\centering
		\includegraphics[width=.49\textwidth]{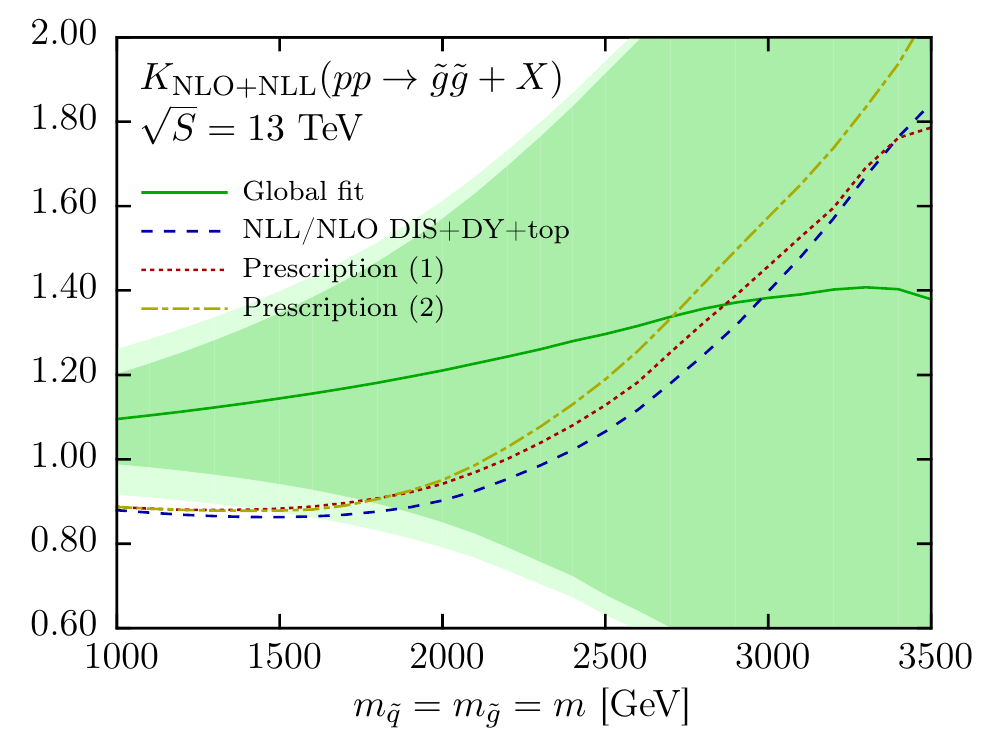}\hfill
                \includegraphics[width=.49\textwidth]{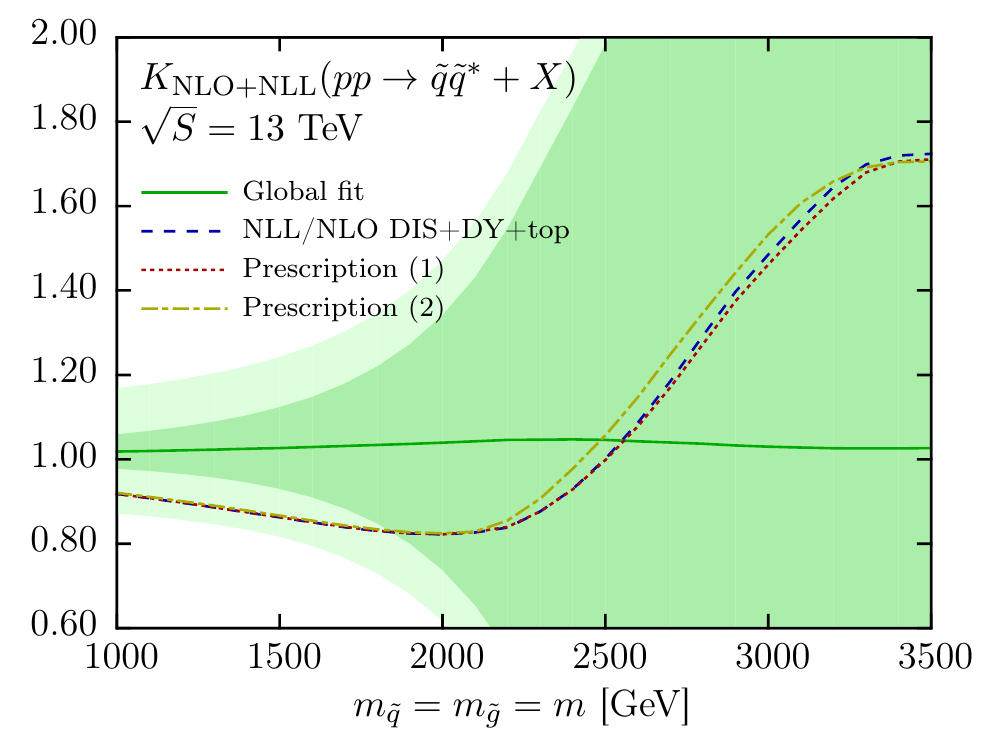}\\
		\includegraphics[width=.49\textwidth]{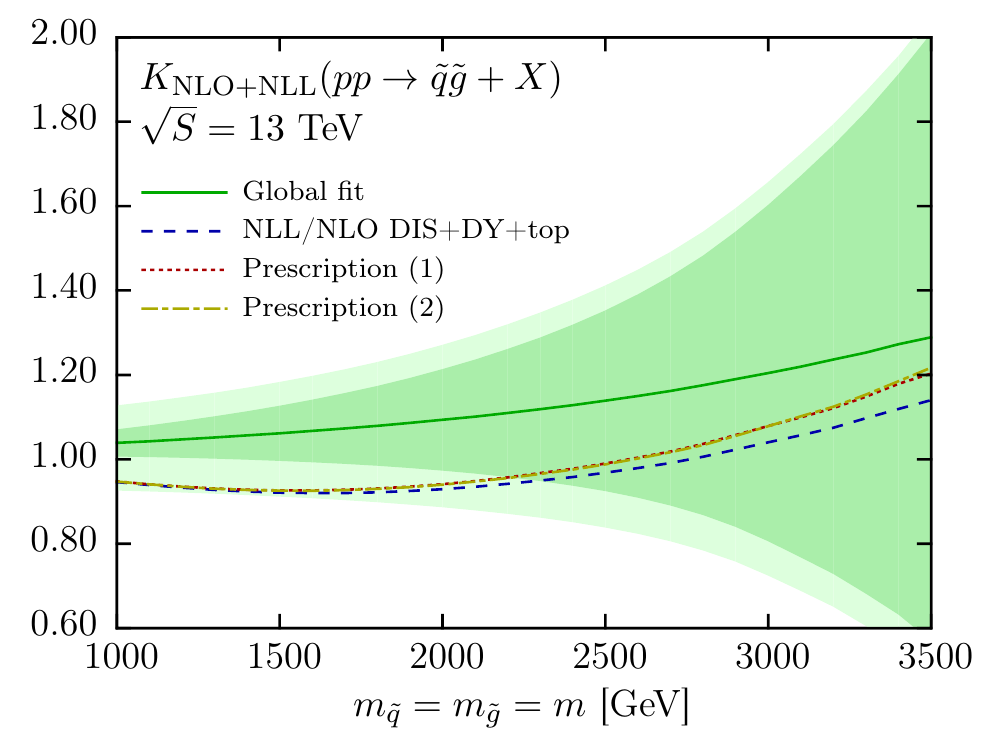} \hfill
                \includegraphics[width=.49\textwidth]{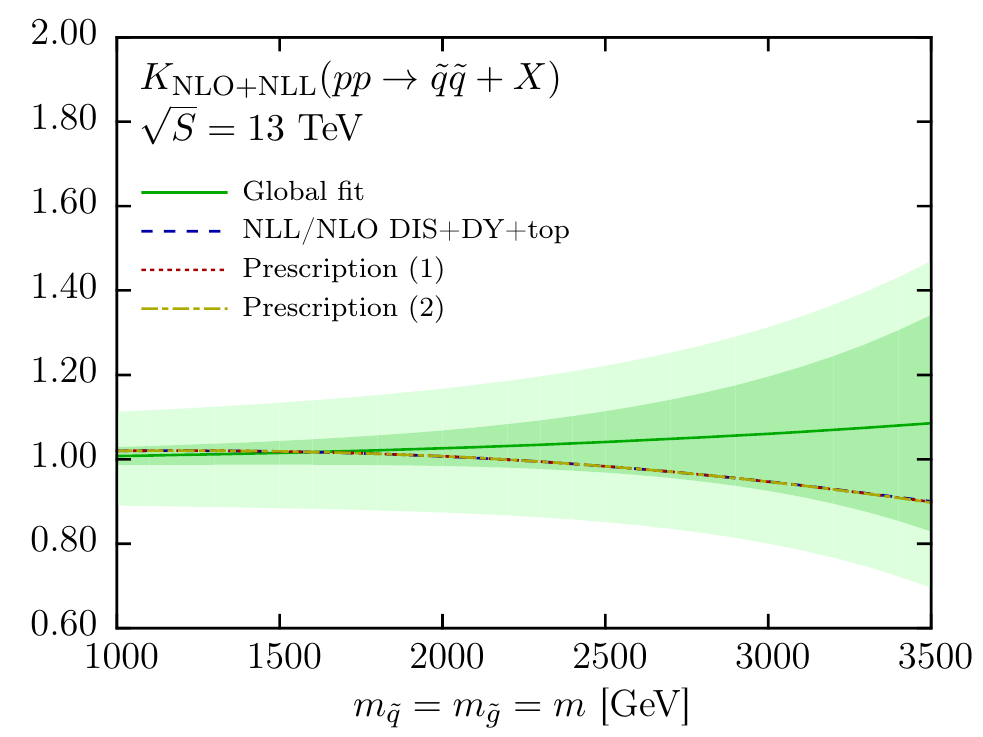}
                \caption{\label{fig:kfact3} \small Comparison of the NLO+NLL $K$-factors
                  obtained using the NNPDF3.0 NLO global fit,
                  Eq.~(\ref{fig:kfact1})  with the corresponding
                  $K$-factors obtained accounting for the effect of resummation in the input
                  PDFs, Eq.~(\ref{eq:prescription1}), called here Prescription (1), and  Eq.~(\ref{eq:prescription2}), called here Prescription (2).
                  In the case of the global fit, we show the total uncertainty band (light green band)
                  as well as the PDF-only uncertainty band (solid green band).
                  We also show the $K$-factor 
                  Eq.~(\ref{eq:kfact2}) determined from the DIS+DY+top fit.
                }
	\end{figure}

Using these two prescriptions, we can now compare 
the results obtained using as an input the global NNPDF3.0 NLO fit, Fig.~\ref{fig:kfact1}, with
those obtained using the threshold-improved PDFs.
This comparison is shown in Fig.~\ref{fig:kfact3}, where we first show
the resummed $K$-factors obtained using NNPDF3.0NLO, Eq.~(\ref{fig:kfact1}),
together with the associated total theory uncertainty band.
For the latter, we separate the PDF-only uncertainty (solid band)
from the total
theory error band (lighter band), where PDF uncertainties
have been added linearly to scale 
uncertainties.
The scale error is estimated by varying simultaneously
the factorization and
renormalization scales up and down by a factor two with respect to their
reference value $\mu_R=\mu_F=m$, the sparticle mass.
We then  show  the corresponding
$K$-factors obtained when
accounting for the effect of resummation in the input
                  PDFs, Eq.~(\ref{eq:prescription1}) and  Eq.~(\ref{eq:prescription2}), using
                  the two different prescriptions for the rescaling.
                  For completeness, we also include the $K$-factor obtained
                  from Eq.~(\ref{eq:kfact2}) determined from the DIS+DY+top fit.

        First of all, we note that the choice of prescription has a very small effect,
        except in the region where the PDF uncertainties are huge.
        For all final states in most
        of the
        relevant sparticle mass region,
        the total theory uncertainty band encompasses the shift
        in the central value induced by the use of resummed PDFs, as compared to the global
        fixed-order fit result.
        The only exception is $\tilde{g}\tilde{g}$ production for $m\le 1.5$ TeV, where the
        agreement is only marginal.
        We also see that as we increase the sparticle mass, PDF uncertainties dominate over
        scale uncertainties (which are roughly independent of $m$), in some cases by a large
        amount.
        For the $\tilde{g}\tilde{g}$, $\tilde{q}\tilde{q}^*$, $\tilde{q}\tilde{q}$ and
        $\tilde{q}\tilde{g}$ final states we find that PDF uncertainties begin to dominate
        over scale uncertainties for $m\ge$ 1, 1.5, 2.5 and 1.5 TeV, respectively.

        It is worth emphasizing that several of the features of
        how the cross-section $K$-factors in Fig.~\ref{fig:kfact3} are
        modified due to the use of resummed PDFs can be directly traced
        back to the corresponding modifications at the level
        of PDF luminosities, as illustrated in
        Fig.~\ref{fig:pdflumi}.
     For instance, the turnover in the $K$-factor for $\tilde{q}\tilde{q}^*$
        production from $K<1$ to $K>1$ around $m\simeq2.5$ TeV can be also
        observed for the central value of the $q\bar{q}$ luminosity.
        This behaviour can, in turn,
        be traced back to two facts. Firstly, as quantified in
        Fig.~\ref{fig:initial_state_splitup}, each of the SUSY final
        states considered here, $\tilde{q}\tilde{q}$, $\tilde{q}\tilde{q}^*$,
        $\tilde{q}\tilde{g}$ and $\tilde{g}\tilde{g}$, has a dominant
        partonic production channel, $qq$, $q\bar{q}$, $qg$ and $gg$
        respectively.
	Secondly, the ratio of the $K$-factors in Eq.~(\ref{eq:kfact2}) and Eq.~(\ref{eq:kfact1}) is approximately given by the ratio of NLO+NLL cross sections computed using resummed and fixed-order DIS+DY+top NNPDFs, while the relation is exact for the ratio of the $K$-factors in Eq.~(\ref{eq:prescription1}) and Eq.~(\ref{eq:kfact1}).
        This also explains why in Fig.~\ref{fig:kfact3} the $K$-factors obtained using prescriptions are very close to $K$-factors constructed from
        Eq.~(\ref{eq:kfact2}). 

Fig.~\ref{fig:kfact3} is the main result of this work: for the first time
        we have performed a NLO+NLL calculation of supersymmetric particle
        pair production at hadron colliders
        accounting for the effects of threshold resummation
        both in the partonic cross-sections and in the PDFs.
        As compared to the results obtained using the global NNPDF3.0NLO fit as input, we find
        that including the effect of resummation in the PDFs modifies
        the resummed NLL $K$-factor
        both in a qualitative and in a quantitative way.
        This shift is however contained within the total theory uncertainty band of the
        NNPDF3.0NLO result, and therefore the use of
        threshold-resummed PDFs does not modify the current SUSY exclusion bounds. 

Similarly to the behaviour of the NLL K-factor, it can be shown that the modification of the NNLL K-factor will be mostly driven by the differences between the NNLL and NNLO PDF luminosities obtained on the basis of  DIS+DY+top fits. Given that the global NNLL K-factors follow the behaviour of the NLL K-factors~\cite{Beenakker:2011gt,Beenakker:2014sma} with the NNLL corrections in general smaller than NLL, and that the impact of threshold resummation in PDF analysis at NNLO appears to be much less than at NLO~\cite{Bonvini:2015ira}, we believe our conclusions regarding the behaviour of the K-factor will not change dramatically after increasing the accuracy to NNLL.

%% file: sec-summary.tex
\section{Summary and outlook}
\label{sec:summary}

In this work we have presented updated NLO+NLL predictions for
squark and gluino pair production at the LHC Run II obtained using the NNPDF3.0
NLO global fit.
Our calculations are based on fixed-order NLO partonic cross-sections
matched to NLL threshold resummation.
We have then studied the impact in the calculation of using 
threshold-improved PDFs together
with the resummed partonic cross-sections, finding
that both the quantitative and qualitative behaviour of the
NLO+NLL cross-sections
is modified.
However, we also find that the shift induced by the resummed PDFs
is contained within the total theory uncertainty band of the standard
calculation.

Given that PDFs with threshold resummation are still  in their infancy, and that
the shift they induce is  within the total theory error band
of the calculation using as input NNPDF3.0 NLO global fit, we prefer to
still adopt the latter in our reference calculations.
This choice is reasonable for the time being
in order to determine exclusion limits from
the searches for supersymmetry at the LHC, though it
would become inadequate in case
of a discovery of supersymmetric particles in the next years.

The main limitation of the NNPDF3.0 threshold resummed sets is the fact that they are based on a reduced dataset.
This forces us to introduce somewhat {\it ad-hoc} prescriptions to combine
them with the global fit results.
In order to improve the situation, and to bypass the need of the prescriptions
(and being able to use the resummed PDFs as central value in our
calculation), it should be important to produce truly global
versions of the resummed fits of Ref.~\cite{Bonvini:2015ira}.
This requires in particular the availability of NLL and NNLL calculations
for inclusive jet production in a format ready to use.

Another important message from our study is the role of
PDF uncertainties in high-mass sparticle pair production.
PDF errors are the dominant source of theoretical uncertainty:
in the case of the discovery of sparticles in the TeV region at
Run II, it would be difficult to accurately pin down their properties
unless one is able to reduce these PDF uncertainties.
Fortunately, it is possible to use the LHC data itself~\cite{Rojo:2015acz}
as input
to the global fit to better constrain the large-$x$ PDFs that drive
the PDF uncertainties in the high-mass region.
Examples of measurements that should be helpful in this respect
are top quark pair production, both for inclusive
cross-sections~\cite{Czakon:2013tha},
and for differential distributions~\cite{Guzzi:2014wia},
high-mass Drell-Yan production~\cite{CMSDY,CMS:2014jea,Aad:2013iua}
and inclusive jet and dijet production~\cite{Chatrchyan:2012bja,Aad:2014vwa}.
For all these processes, measurements from ATLAS, CMS and LHCb at Run I
are already available, and more precise data from Run II
will soon extend their kinematical coverage well into the TeV region.

In addition to the phenomenology, it should be emphasized that NLO
supersymmetric
pair production provides a highly non-trivial theoretical laboratory
to test the perturbative convergence of perturbative QCD calculations.
In this respect, the availability for the first time of threshold-improved
PDFs provides a unique opportunity to test in detail the interplay
of the effects of higher-order resummation in the various pieces
of the calculation.

The updated NLO+NLL squark and gluino production cross-sections
with the NNPDF3.0NLO global fit are available in the
{\tt NLL-fast} format~\cite{NNLfastWebpage}.
In addition, the cross-sections obtained in this work
using the threshold-resummed
PDFs are also available from the authors upon request.

\subsection*{Acknowledgments}

We are grateful to Marco Bonvini, Luca Rottoli and Maria Ubiali
for useful discussions on threshold-resummed PDFs and for
their collaboration in early stages of this work.
W.~B. and E.~L  have been supported by the
Netherlands Foundation for Fundamental
Research of Matter (FOM) programme 156, entitled  ``Higgs as Probe and Portal'',
and the National Organization for Scientific Research (NWO).
This work was also supported by the Research Executive Agency (REA)
of the European Union under Grant Agreement number
PITN-GA2012-316704 (HiggsTools).
M.~K. is supported in part by the DFG research unit 
``New physics at the LHC".
The work of A.~K. is partially supported by the DFG grant KU 3103/1. C.B. is supported by BMBF Verbundprojekt 05H2015 (BMBF-FSP 104).
The work of S.~M. is supported by the U.S.\ National Science Foundation, under grant PHY--0969510, the LHC Theory Initiative.
J.~R. is supported by an STFC Rutherford Fellowship
and Grant ST/K005227/1 and ST/M003787/1 and
by an European Research Council Starting Grant ``PDF4BSM".
%


%% file: sec-appendix.tex
\section{Threshold-resummed PDFs from a scheme transformation}
\label{sec:tresh-resumm-pdfs}

The threshold-improved PDFs of Ref.~\cite{Bonvini:2015ira} 
incorporate, though the global fitting procedure,
the large logarithms of the cross-sections involved in the fit to all orders.
Such logarithms are universal at the leading logarithmic level, and 
to a considerable extent at the next-to-leading level and beyond as well.
This incorporation is then numerical in nature, and indeed, as 
we observe in this paper, corrections for other
cross-sections are reduced, because a large part of the resummed threshold
logarithms in the higher corrections now reside in the PDFs.
In this appendix we point out another method by which one 
might include large threshold logarithms in the PDFs, using a dedicated
mass factorization scheme choice.

In general one may represent mass factorization of initial state collinear
divergences for PDFs in the following schematic way
\begin{equation}
  \label{eq:a1}
  \hat{f}_i =  \Gamma_{ik,\mathrm{scheme}}
  (\alpha_s(\mu),Q^2,\mu^2,\varepsilon)\otimes f_k(\mu) \, ,
\end{equation}
where $i,k$ label parton flavours, $\hat{f}_i$ ($f_i$) are 
bare (factorized) PDFs, $\mu$ is the factorization scale,
and $Q$ a representative physical scale for the process.
The functions
$\Gamma_{ik,\mathrm{scheme}}$ are known as
\emph{transition
  functions}, and take the general form
\bea
  \label{eq:a2}
 &&\Gamma_{ik,\mathrm{scheme}}  (\alpha_s(\mu),Q^2,\mu^2,\varepsilon)(z) 
 = \delta(1-z)\delta_{ik} + \\ \nonumber
&& \frac{\alpha_s(\mu)}{8\pi}\left(\frac{-1}{\varepsilon} +\gamma_E
   - \ln 4\pi  \right) P_{ik}(z) +\ldots + f_{ik} (\alpha_s(\mu),Q^2,\mu^2,z) \,,
\eea
where the ellipses indicate higher order pole terms, and the
expression is in $4-2\varepsilon$ dimensions.
The choice of finite terms $f_{ik}$ then determines the
factorization
scheme.
In Eq.~(\ref{eq:a2}) $ P_{ik}(z)$ are the standard DGLAP
splitting functions.

In the following,
we set
$\mu^2=Q^2$ in these functions, and drop the corresponding arguments.
One may 
choose the $f_{ik}$ according to specific preference,
provided that the
charge and momentum sum rules for the PDFs so defined remain
satisfied. For instance, for the momentum sum rule the following conditions must hold 
\begin{align}
\label{eq:a4}
&   \int dz\, z\, \left(\Gamma_{gg} + 2n_f \Gamma_{qg}   \right)=1 \,,  \nonumber \\
&   \int dz\, z\, \left(\Gamma^S_{qq} + \Gamma_{gq}  \right) =1\,,
\end{align}
where the superscript $S$ indicated the
singlet combination of quark PDFs (sum over
all active quarks and antiquarks), and $n_f$ is the
number of active quark flavours.

All modern PDF sets are
defined in the $\overline{\mathrm{MS}}$ scheme, for which  one has
\begin{equation}
  \label{eq:a3}
  f_{ik} = 0\,.
\end{equation}
The advantages of this scheme are first
process independence, and second, ease of implementation.
Note that due to the properties of the splitting functions the charge and momentum sum
rules (\ref{eq:a4}) are automatically preserved.
In the past also the DIS scheme has been used. In this 
scheme one chooses the $f_{ik}$ functions such that the non-singlet
DIS quark coefficient is equal to its lowest order
expression, to any order
in perturbation theory.
Besides the DIS scheme, also a Drell-Yan scheme has been
proposed in~\cite{Smith:1989xz}, for the computation of
inclusive $W+\gamma$ production at NLO. 

One could also define a variant of this
scheme, the threshold-resummation (TR) scheme, 
in which {\it e.g.} the Drell-Yan and Higgs cross-sections are
defined to be
equal to their lowest order cross-sections, and the $qg$ and
$\bar{q}g$ transition functions are defined such that the sum rules
Eq.~(\ref{eq:a4})
are satisfied.
In such a TR scheme, when fitting to data, 
the threshold logarithms are largely subtracted from the partonic 
cross-sections in the fit, by design, and absorbed into the TR scheme PDFs.
To use such TR scheme PDFs for other partonic cross-sections one must 
of course subtract the $f_{ik}$ functions  for those cross-sections
as well.
If these cross-sections are also threshold resummed, a large amount of
cancellation should occur through that subtraction. 

Thus, for the TR scheme one might choose the Drell-Yan and the Higgs
cross section
\begin{equation}
  \label{eq:a5}
  f_{qq}(z) = f_{\bar{q}\bar{q}}(z)= \frac{1}{2} \sigma_{q\bar{q}}^{\mathrm{DY,N^iLL}}(z)\,,
\end{equation}
and 
\begin{equation}
  \label{eq:a6}
  f_{gg}(z) = \frac{1}{2} \sigma_{gg}^{H,\mathrm{N^iLL}}(z)\,.
\end{equation}
For the latter in particular one might also consider the inclusive top
cross section, which is dominated by the gluon distribution as well,
but its colour structure is more involved than Higgs production. 
Whether one takes on the right hand side of
Eqs.~(\ref{eq:a5}, \ref{eq:a6}) only the logarithmic terms or uses the full matched
resummed cross section, including hard part, is a matter of choice.

The advantage of such a TR factorization scheme 
would be that it is analytical, allowing for better insight into its
effects.  The downside however is clearly that substantial
additional work would be
required for users of such TR scheme PDFs, who must convert their
partonic cross-sections to this scheme.
Moreover, a TR scheme method would be more constraining because the 
sum rules (\ref{eq:a4}) must be obeyed, so any changes in subtraction
terms are highly constrained. For instance, the off-diagonal
transition functions $\Gamma_{qg}$ and $\Gamma_{gq}$ also
incorporate leading logarithmic effects, via the sum rules.

In the approach of \cite{Bonvini:2015ira}  compliance of
Eq.~(\ref{eq:a4})
is ensured in the fitting procedure, and holds, whatever
cross-sections one chooses to include in the fit.
Therefore, while a more extended study of such a TR scheme approach
might be interesting, employing the framework of~\cite{Bonvini:2015ira} 
in this paper is advantageous since
it achieves the same goals with greater flexibility.

%% file: SUSYnnpdf.bbl
\providecommand{\href}[2]{#2}\begingroup\raggedright\endgroup